\documentclass[aps,pra,twocolumn,floatfix,superscriptaddress]{revtex4}

\usepackage{amssymb,amsmath,amstext}                
\usepackage{graphicx}                                               
\usepackage{epstopdf}                                               
\usepackage{color}                                                     
\usepackage{bm}                                                        
\usepackage{appendix}                                              
\usepackage[utf8]{inputenc}
\usepackage{bbold}
\usepackage{bbm}
\usepackage{ulem}
\usepackage{latexsym}
\usepackage[colorlinks=true,citecolor=blue,linkcolor=magenta]{hyperref}

\newcommand{\vect}[1]{\mathbf{#1}}

\def\be{\begin{equation}}
\def\ee{\end{equation}}
\def\bea{\begin{eqnarray}}
\def\eea{\end{eqnarray}}
\def\ra{\rangle}
\def\la{\langle}
\def\bi{\begin{itemize}}
\def\ei{\end{itemize}}
\def\ben{\begin{enumerate}}
\def\een{\end{enumerate}}

\definecolor{dgreen} {RGB}{78,138,21}

\begin{document} 

\title{Time crystals: analysis of experimental conditions}

\author{Krzysztof Giergiel} 
\affiliation{
Instytut Fizyki imienia Mariana Smoluchowskiego, 
Uniwersytet Jagiello\'nski, ulica Profesora Stanis\l{}awa \L{}ojasiewicza 11, PL-30-348 Krak\'ow, Poland}
\author{Arkadiusz Kosior} 
\affiliation{
Instytut Fizyki imienia Mariana Smoluchowskiego, 
Uniwersytet Jagiello\'nski, ulica Profesora Stanis\l{}awa \L{}ojasiewicza 11, PL-30-348 Krak\'ow, Poland}
\author{Peter Hannaford}
\affiliation{Centre for Quantum and Optical Science, Swinburne University of Technology, Hawthorn, Victoria 3122, Australia}
\author{Krzysztof Sacha} 
\affiliation{
Instytut Fizyki imienia Mariana Smoluchowskiego, 
Uniwersytet Jagiello\'nski, ulica Profesora Stanis\l{}awa \L{}ojasiewicza 11, PL-30-348 Krak\'ow, Poland}
\affiliation{Mark Kac Complex Systems Research Center, Uniwersytet Jagiello\'nski, ulica Profesora Stanis\l{}awa \L{}ojasiewicza 11, PL-30-348 Krak\'ow, Poland
}

\begin{abstract}
Time crystals are quantum many-body systems which are able to self-organize their motion in a periodic way in time. Discrete time crystals have been experimentally demonstrated in spin systems. However, the first idea of spontaneous breaking of discrete time translation symmetry, in ultra-cold atoms bouncing on an oscillating mirror, still awaits experimental demonstration. Here, we perform a detailed analysis of the experimental conditions needed for the realization of such a discrete time crystal. Importantly, the considered system allows for the realization of dramatic breaking of discrete time translation symmetry where a symmetry broken state evolves with a period tens of times longer than the driving period. Moreover, atoms bouncing on an oscillating mirror constitute a suitable system for the realization of dynamical quantum phase transitions in discrete time crystals and for the demonstration of various non-trivial condensed matter phenomena in the time domain. We show that Anderson localization effects, which are typically associated with spatial disorder and exponential localization of eigenstates of a particle in  configuration space, can be observed in the time domain when ultra-cold atoms are bouncing on a randomly moving mirror.
\end{abstract}
\date{\today}

\maketitle

\section{Introduction}
\label{intro}

Time crystals are quantum many-body systems which due to interactions between the particles are able to self-organize in a periodic way in time which is in full analogy to the formation of space crystals due to mutual interactions between atoms in condensed matter physics \cite{Sacha2017rev}. Frank Wilczek initiated time crystal research but his original idea concerning the formation of a crystalline structure in time turned out to be impossible to realize because he considered a time-independent system in the ground state \cite{Wilczek2012,Bruno2013b,Watanabe2015,Syrwid2017,Iemini2017,Huang2017a,Prokofev2017}. However, soon after, another version of time crystals was proposed in which a periodically driven quantum many-body system spontaneously breaks discrete time translation symmetry of the Hamiltonian and starts evolving with a period twice as long as the period of the external driving \cite{Sacha2015,Khemani16,ElseFTC,Yao2017,Lazarides2017,Russomanno2017,Zeng2017,Nakatsugawa2017,Ho2017,Huang2017,Gong2017,Wang2017,Mizuta2018}. This kind of quantum self-reorganization of the motion of quantum many-body systems, called discrete time crystals, has already been realized experimentally in spin systems \cite{Zhang2017,Choi2017,Nayak2017,Pal2018,Rovny2018,Rovny2018a,Autti2018}. It should be mentioned that in the classical regime breaking of discrete time translation symmetry in an atomic system has also been demonstrated in the laboratory \cite{Kim2006,Heo2010}.  

In the present paper we return to the first experimental proposal of a discrete time crystal and analyze in detail the conditions for its experimental realization. In Ref.~\cite{Sacha2015} it was shown that an ultra-cold atomic cloud bouncing on an oscillating atom mirror is able to spontaneously self-reorganize its motion and to move with a period twice as long as the mirror oscillation period if the interactions between atoms are sufficiently strong. The system itself is suitable for realization not only of spontaneous breaking of discrete time translation symmetry of the Hamiltonian into a motion with twice as long a period but also into a motion with any multiple period of the driving. We focus here on the self-reorganization of the motion of the system with a period 40 times longer than the period of the driving. Such a dramatic breaking of discrete time translation symmetry of the Hamiltonian is not possible to observe in spin systems unless one deals with a very large spin quantum number.  

The formation of a discrete time crystal is related to a quantum phase transition \cite{Sacha2015,Zin2008}. In order to form a time crystal, the strength of the interactions between atoms bouncing on a mirror must be greater than a critical value. It turns out that when one suddenly changes the interaction strength from the time crystal regime to the non-interacting regime, a dynamical quantum phase transition can be observed in the time evolution of the system \cite{Kosior2017}. Dynamical quantum phase transitions are recently discovered analogs of equilibrium phase transitions where the non-analytical behavior of a system is observed not as a function of a control parameter but versus time \cite{Heyl2013,Jurcevic2017,Flaschner2018,Heyl2018rev}. We analyze the experimental conditions which allow one to observe that the return probability of the system to the initial degenerate manifold reveals a cusp at a critical moment of time after the sudden change from the time crystal phase to the non-interacting regime.


Ultra-cold atoms bouncing on an oscillating atom mirror \cite{Steane95} (for stationary mirror experiments see \cite{Roach1995,Sidorov1996,Westbrook1998,Lau1999,Bongs1999,Sidorov2002,Fiutowski2013,Kawalec2014})
constitute a promising system for experimental realization of various condensed matter phases in the time domain \cite{Sacha2017rev}. If a single- or many-body system is periodically and resonantly driven and the resonance is of a high order, such as 40 times longer than the driving period, the system behaves like electrons in a space crystal \cite{Guo2013,Sacha15a}. Importantly, such a crystalline behavior of a resonantly driven system is observed in the time domain when the detection is carried out in the laboratory frame. A proper manipulation of the periodic driving allows one to realize various solid state phenomena in the time domain \cite{Giergiel2018}. In the present paper we focus on Anderson localization in time \cite{Sacha15a,sacha16,Giergiel2017,delande17} and show that the introduction of randomness in the motion of the atom mirror leads to the Anderson localization phenomenon which is observed in the laboratory frame as an exponential localization versus time of the probability for the detection of atoms at a fixed position. 

The paper is organized as follows. In Sec.~\ref{system} we introduce the system and provide the theoretical description needed to analyze the phenomena we are interested in. In Sec.~\ref{spont_DTC} we perform a detailed analysis of the experimental conditions required for the realization of discrete time crystals. The results of Sec.~\ref{spont_DTC} also serve as a base for an analysis of the experimental conditions for realization of dynamical quantum phase transitions in time crystals, Sec.~\ref{DQPT}, and Anderson localization in time, Sec.~\ref{AL_in_time}. We conclude in Sec.~\ref{concl}.

\section{Description of the system}
\label{system}

In this section we introduce the system and its theoretical description. We begin with a single-particle problem which is followed by the many-body generalization.  

\subsection{Single-particle problem}
\label{system_singl}

In the present paper we use the gravitational units where the length, energy and time are given by
\bea
l_0=\left(\frac{\hbar^2}{m^2g}\right)^{1/3}, \quad E_0=mgl_0, \quad t_0=\frac{\hbar}{mgl_0},
\label{units}
\eea
respectively, with $m$ the atom mass and $g$ the gravitational acceleration. In the laboratory frame a single atom bouncing on an oscillating atom mirror is described by the following Hamiltonian
\be
H=\frac{p^2}{2}+z+F\left(z-f(t)\right),
\label{h_lab}
\ee
where $F(z)$ describes the mirror, i.e., the profile of the reflecting potential, and $f(t)=f(t+T)$, with 
\be
T=\frac{2\pi}{\omega},
\ee
determines the periodic oscillations of frequency $\omega$ of the mirror position. A theoretical description of the system is much more convenient when we switch from the laboratory frame to the frame oscillating with the mirror. Then, the mirror is fixed but the effective  gravitational acceleration oscillates in time
\be
H=\frac{p^2}{2}+z+zf''(t)+F\left(z\right).
\label{h_v1}
\ee
Except in Sec.~\ref{AL_in_time}, we focus on the case where 
\be
f(t)=-\gamma \cos\omega t, \quad \gamma=\frac{\lambda}{\omega^2}.
\label{mirror}
\ee 
The ratio $\lambda/\omega^2$ determines the amplitude of the harmonic oscillations of the mirror in the laboratory frame.
Assuming   the mirror can be modeled by a hard wall potential located at $z=0$ in the oscillating frame, we may drop the $F(z)$ in Eq.~(\ref{h_v1}) which leads to the final form of the single particle Hamiltonian \cite{Buchleitner2002}

\be
H(t)=\frac{p^2}{2}+z+\lambda z\cos\omega t, \quad z\ge 0.
\label{h}
\ee

The energy of the considered system is not conserved because the Hamiltonian (\ref{h}) depends explicitly on time. However, due to the time periodicity, there are eigenstates of the so-called Floquet Hamiltonian \cite{Shirley1965,Buchleitner2002}, 
\be
{\cal H}=H(t)-i\partial_t, 
\label{floqh}
\ee
which evolve periodically with the period of the driving. The corresponding eigenvalues are called quasi-energies of the system. The Floquet formalism is in full analogy to the Bloch theorem approach known in condensed matter physics. The aim of the present paper is to analyze experimental conditions of the Floquet eigenstates of the single-particle system (\ref{h}) and its many-body version. 

The description of a particle bouncing resonantly on an oscillating mirror, which we are interested in, can be simplified by employing the quantum secular approximation \cite{Berman1977}. However, starting with the classical description, which at the end is quantized, we not only arrive at the same results but also gain an intuitive picture of the system dynamics. 

Let us begin with the classical version of the Hamiltonian (\ref{h}) and apply a canonical transformation from the Cartesian position and momentum to the so-called action-angle variables $I$ and $\theta$ of the unperturbed Hamiltonian $H_0=p^2/2+z$ \cite{Buchleitner2002,Lichtenberg1992}. In terms of this new pair of  canonically conjugate variables, the unperturbed Hamiltonian depends on the momentum (action) only, $H_0(I)=(3\pi I)^{2/3}/2$, and the periodic particle trajectory can be found immediately. Indeed, the action $I=\rm constant$ and the position of the particle on a periodic orbit is given by an angle which changes linearly in time, $\theta(t)=\Omega(I)t+\theta(0)$, where $\Omega(I)=dH_0(I)/dI$ is the frequency of the periodic motion of an unperturbed particle. The distance of the classical turning point of a particle from the mirror is 
\be
h=\frac{\pi^2}{2\Omega^2}. 
\ee
In the presence of the mirror oscillations, we are interested in the motion of a particle in the vicinity of a periodic orbit which is resonant with the time-dependent driving, i.e.,  when $I\approx I_s$, where $\omega=s\Omega(I_s)$ with integer $s$. The $s:1$ resonant motion can be described by a simple effective Hamiltonian when we switch to the frame moving along a resonant orbit, defined by 
\be
\Theta=\theta-\frac{\omega}{s}t, 
\label{movlab}
\ee 
and perform averaging over time. In the moving frame, $\Theta$ and $P=I-I_s$ are slow variables provided we are close to a resonant orbit, i.e., provided $P\approx 0$. Then, averaging the Hamiltonian (\ref{h}) over time yields \cite{Buchleitner2002,Lichtenberg1992}
\be
H_{\rm eff}\approx \frac{P^2}{2m_{\rm eff}}+V_0\cos(s\Theta),
\label{heff}
\ee
where a constant term has been omitted and 
\be
m_{\rm eff}=-\frac{\pi^2s^4}{\omega^4}, \quad V_0=\frac{\lambda}{\omega^2}(-1)^{s+1}.
\ee 

The Hamiltonian (\ref{heff}) indicates that in the moving frame, a resonantly bouncing atom behaves effectively like a particle in a periodic lattice, i.e., like an electron in a space crystal \cite{Guo2013,Sacha15a,Guo2016,Guo2016a,Liang2017}. Importantly, the crystal behavior in the moving frame will be reproduced in the time domain when we return to the laboratory frame \cite{Sacha15a,sacha16,Mierzejewski2017,Giergiel2018}. Indeed, the linear relation (\ref{movlab}) between the position $\Theta$ in the moving frame and the time $t$ in the laboratory frame ensures that in the quantum description when we fix the position in the laboratory frame $\theta=$const., the crystalline behavior in $\Theta$, described by the effective Hamiltonian (\ref{heff}), will be observed in the evolution of the probability of detection of an atom versus $t$. This simple argumentation shows that it is not the presence of any external space periodic potential but the resonant dynamics itself that is responsible for the emergence of a crystalline behavior. 
It also shows that we have  a platform for realization of a broad range of condensed matter phenomena in the time domain --- as an example we analyze Anderson localization in time in Sec.~\ref{AL_in_time}. However, in Sec.~\ref{spont_DTC} and \ref{DQPT} we consider a different aspect of the system. We show that ultra-cold atoms bouncing resonantly on a harmonically oscillating mirror can reveal spontaneous breaking of the discrete time translation symmetry of the Hamiltonian if the interaction between atoms is sufficiently strong and a discrete time crystal phenomenon forms \cite{Sacha2015,Kosior2017}.

\begin{figure} 	            
\includegraphics[width=1.\columnwidth]{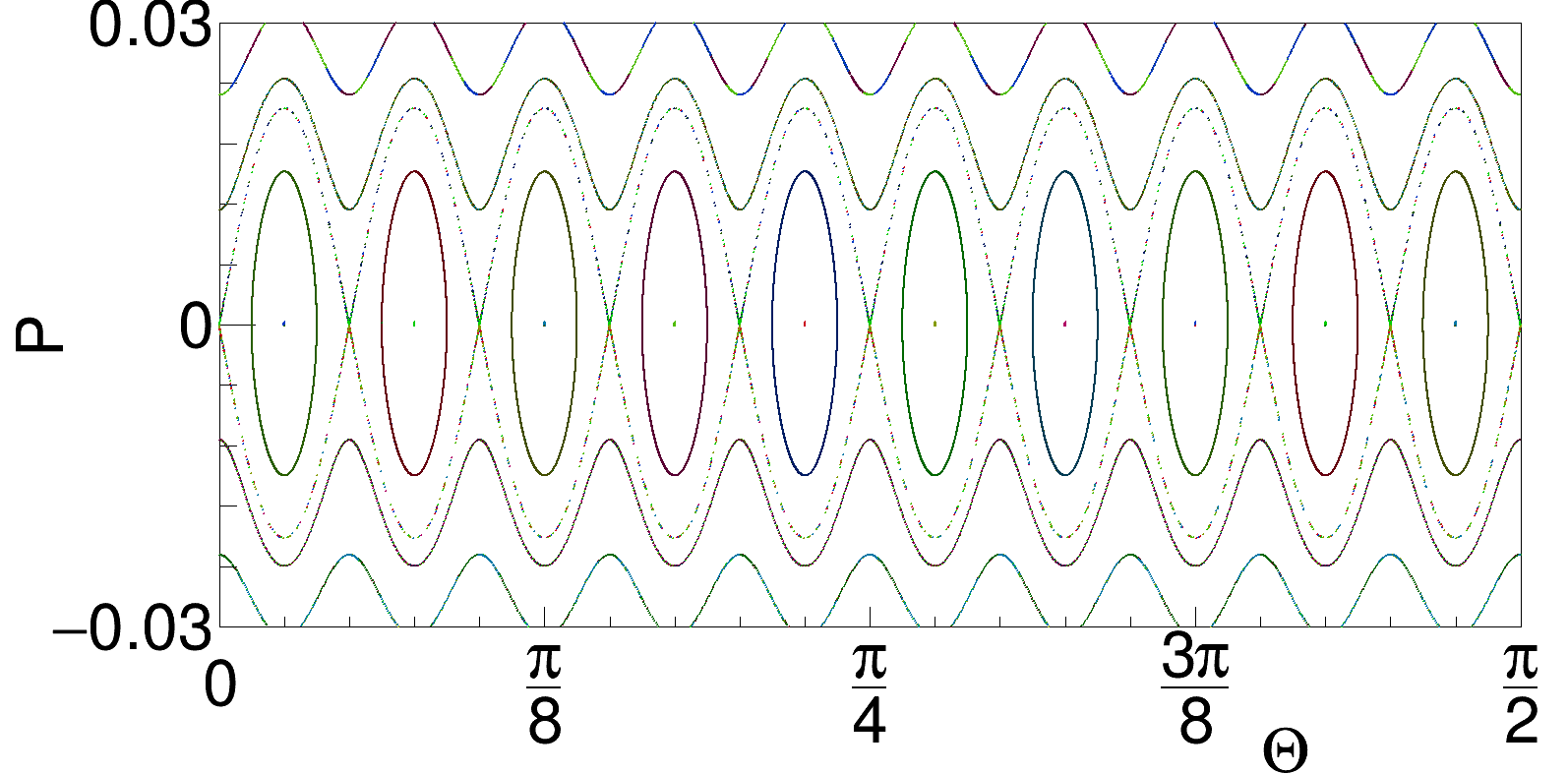}       
\includegraphics[width=1.\columnwidth]{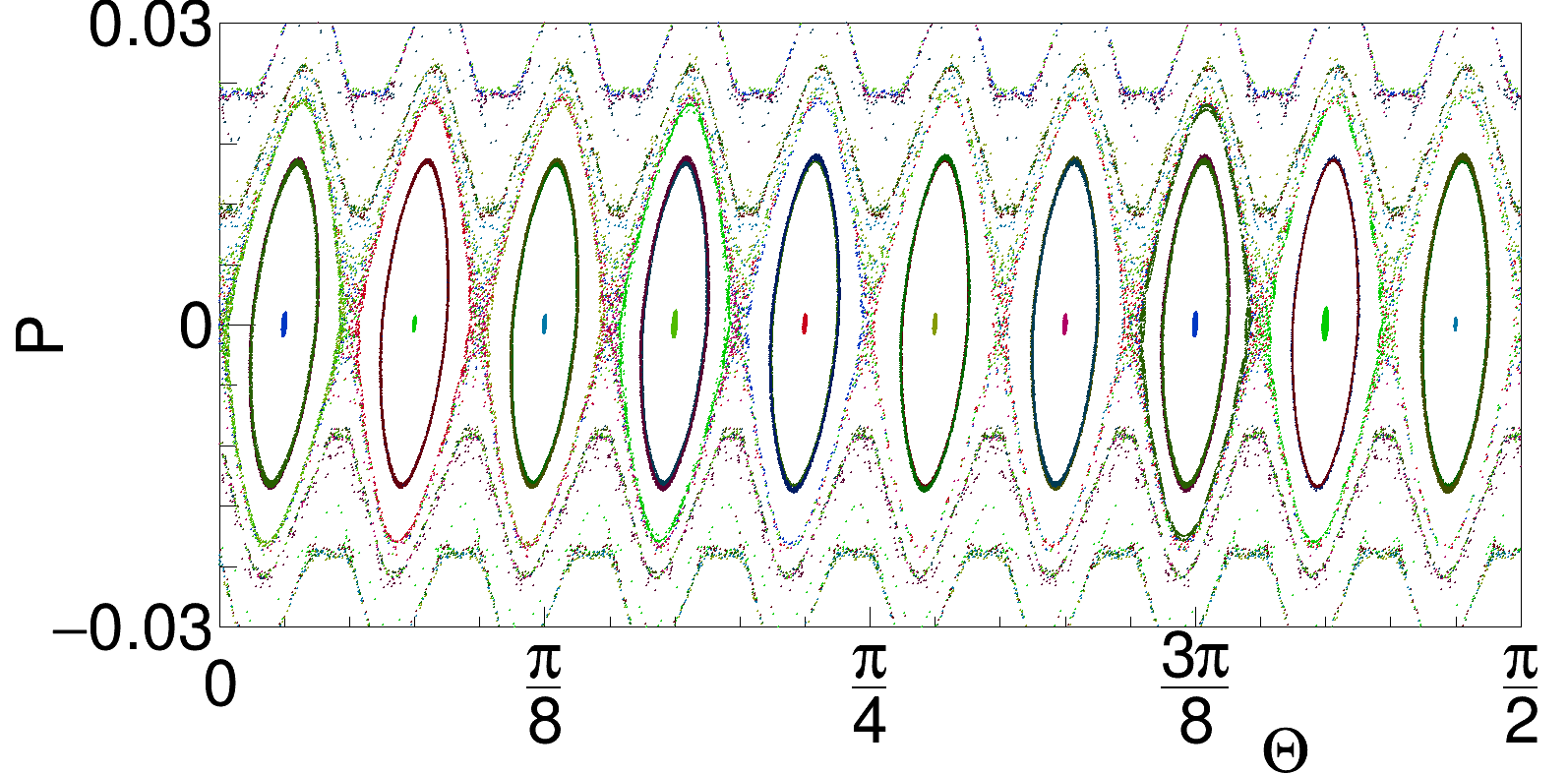}       
\includegraphics[width=1.\columnwidth]{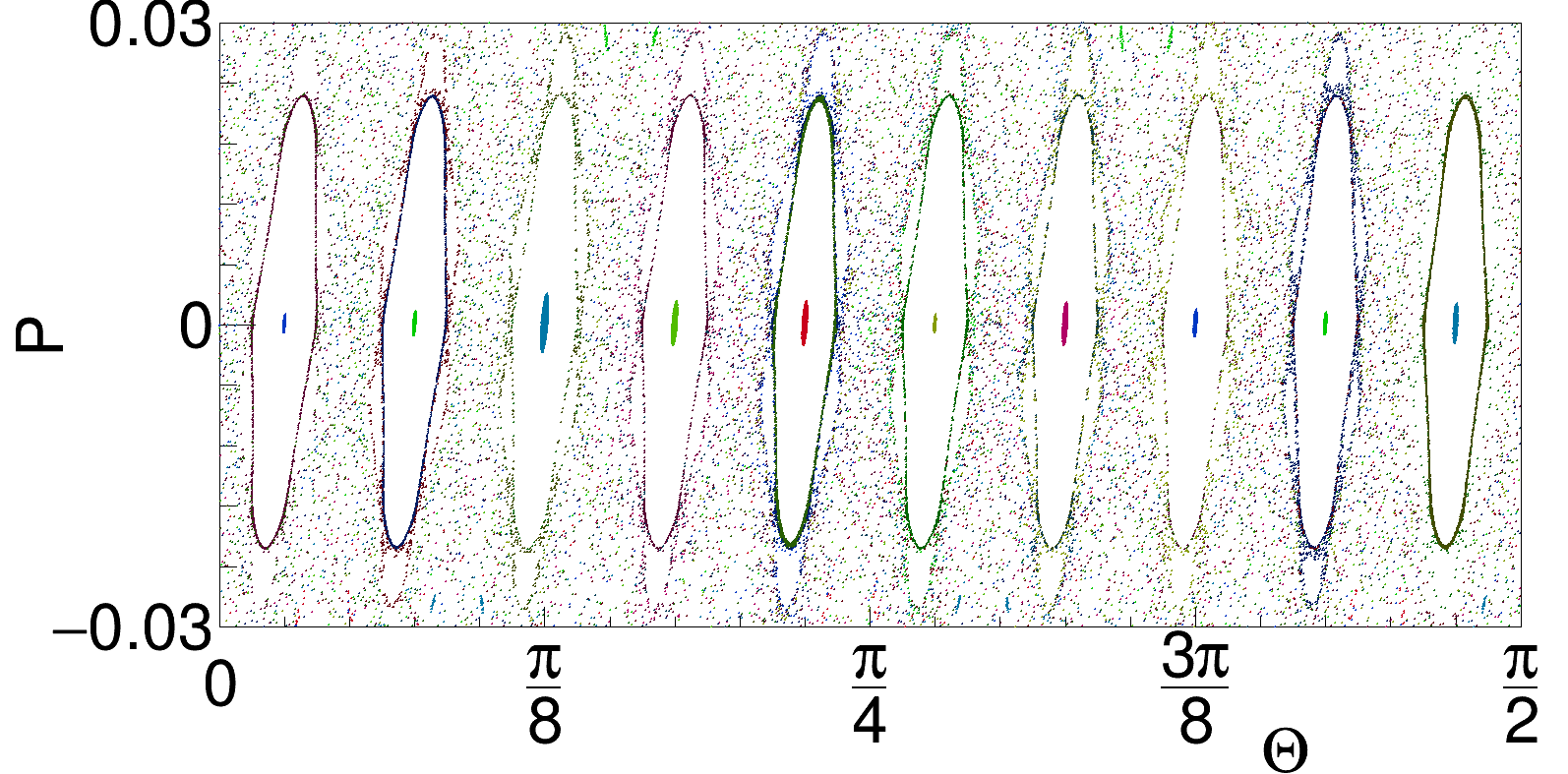}       
\caption{Top panel:   phase space portrait generated by the effective Hamiltonian (\ref{heff}) for $\lambda=0.2$. Middle and bottom panels show the stroboscopic phase space portraits generated by the exact Hamiltonian (\ref{h}) for $\lambda=0.2$ and $\lambda=0.4$, respectively. The frequency of the mirror oscillations, $\omega=40(\pi^2/3)^{1/3}$,  corresponds to the $40:1$ resonance ($s=40$) for the resonant value of the action $I_{40}=1$. Note that for the sake of clarity we show only a quarter of the full $2\pi$ range of $\Theta$.} 
\label{fig1}   
\end{figure}

Before we switch to the description of time crystals we need to analyze the validity of the approach we use. Classical equations of motion generated by the Hamiltonian (\ref{h}) possess scaling symmetry \cite{Buchleitner2002}. It means that when we appropriately rescale the parameters and the position and momentum, the equations of motion do not change. It allows us to set one of the parameters, perform the analysis of the system and use the results for other values of the chosen parameter by applying the scaling transformation. Let us set the resonant action $I_s=1$ which implies the resonant driving frequency $\omega=s(\pi^2/3)^{1/3}$ and perform an analysis of the validity of the effective Hamiltonian (\ref{heff}) for different values of the perturbation amplitude $\lambda$. The obtained results will allow us to predict the resonant behavior for any value of $I_s$ provided they are rescaled according to $\omega'=I_s^{-1/3}\omega$, $\lambda'=\lambda$, $p'=I_s^{1/3}p$, $x'=I_s^{2/3}x$ and $t'=I_s^{1/3}t$. In Fig.~\ref{fig1} phase space portraits corresponding to the $40:1$ resonance ($s=40$), $\lambda=0.2$ and $\lambda=0.4$ are presented. In the latter case the phase space possesses a significant chaotic area but in the former case the phase space is mostly regular and resembles the behavior predicted by the effective Hamiltonian (\ref{heff}).

In the quantum description the scaling symmetry is broken because the Planck constant sets a scale in   phase space which can be easily seen from the commutation relation
\be
[x',p']=i \quad \Rightarrow \quad [x,p]=\frac{i}{I_s},
\ee
which also indicates that $I_s^{-1}$ can be treated as an effective Planck constant. 
For $I_s\gg 1$, the quantized version of the classical effective Hamiltonian (\ref{heff}), i.e., when $P\rightarrow -i\frac{\partial}{\partial \Theta}$, provides a perfect quantum description of the resonant behavior of the system \cite{Buchleitner2002}. As already mentioned the same quantum results can be obtained by applying the quantum secular approximation \cite{Berman1977}. 

Now we can define a simple strategy on how to choose suitable parameters of the system. If we are interested in a certain $s:1$ resonance behavior, we set $I_s=1$ and $\omega=s(\pi^2/3)^{1/3}$ and perform an analysis of which values of $\lambda$ are allowed in order to deal with the behavior predicted by the effective Hamiltonian (\ref{heff}). Then, in order to choose which value of $I_s$ is suitable in the quantum case, we have to keep in mind that the effective Planck constant $I_s^{-1}$ must be smaller or at least comparable to the area of a single regular elliptic island (cf. Fig.~\ref{fig1}), otherwise no quantum state is localized (in a semiclassical sense) in an island and the picture of a particle moving like an electron in a space crystal is lost. Having determined $I_s$ and employing the scaling transformation we obtain the desired $\omega'=I_s^{-1/3}s(\pi^2/3)^{1/3}$ and $\lambda'=\lambda$.

Eigenstates of the effective Hamiltonian (\ref{heff}) obtained in the moving frame correspond to the Floquet states of the original Floquet Hamiltonian (\ref{floqh}). For $s\gg 1$ the eigenstates of (\ref{heff}) are Bloch waves \cite{Giergiel2018}. If we are interested in the first energy band of (\ref{heff}) only, the effective description can be further  simplified. That is, expanding the wave function of the quantum version of (\ref{heff}) in the basis of  Wannier states $w_i(\Theta)$ localized in each site of the effective periodic potential in (\ref{heff}) and belonging to the first energy band, $\psi(\Theta)=\sum_{i=1}^sa_iw_i(\Theta)$, we obtain an expression for the particle energy in the tight binding form \cite{Dutta2015}
\be
E\approx-\frac{J}{2}\sum\limits_{i=1}^s\left(a_{i+1}^*a_i+c.c.\right),
\label{1dtight}
\ee
where $J=-2\la w_{i+1}|H_{\rm eff}|w_i\ra$ is the amplitude of tunneling of an atom between neighboring Wannier states. Note that the Wannier states are represented in the laboratory frame by localized wavepackets evolving periodically along the $s:1$ resonant orbit with a period $s$ times longer than the mirror oscillation period $T$.

\subsection{Many-body problem}

The interaction between atoms couples different degrees of freedom and we may not consider different degrees of freedom independently without justification. 

We assume that a cloud of ultra-cold atoms is bouncing on the oscillating atom mirror in the presence of a transverse harmonic potential characterized by the frequency $\omega_\perp$. The many-body Hamiltonian of our system, in the second quantization formalism and in the gravitational units (\ref{units}), reads \cite{Pethick2002}
\bea
\hat H&=&\int d^3r\;\hat\Psi^\dagger\left[\frac{\vect p^2}{2}+z+\lambda z\cos\omega t+\frac{\omega_\perp^2(x^2+y^2)}{2}\right.
\cr 
&& \left.+\frac{g_0}{2}\hat\Psi^\dagger\hat\Psi\right]\hat\Psi, 
\label{hath3d}
\eea 
where $\hat \Psi(\vect r,t)$ is the bosonic field operator and the contact interaction between atoms is described by the coefficient $g_0=4\pi a_s$, with $a_s$ being the atomic s-wave scattering length given in the gravitational units (\ref{units}).
If the transverse confinement is sufficiently strong (see the analysis in Sec.~\ref{DTCtheory}), ultra-cold atoms do not populate excited states along the transverse directions and we may restrict to the ground state $\phi_0(x,y)$ of the harmonic oscillator trap of   frequency $\omega_\perp$. That is, substituting $\hat\Psi(\vect r,t)=\phi_0(x,y)\hat\psi(z,t)$ in (\ref{hath3d}) and integrating over $x$ and $y$ we arrive at the one-dimensional version of the many-body Hamiltonian \cite{Pethick2002}
\be
\hat H=\int\limits_0^\infty dz\;\hat\psi^\dagger\left[\frac{p^2}{2}+z+\lambda z\cos\omega t+\frac{g_{\rm 1D}}{2}\hat\psi^\dagger\hat\psi\right]\hat\psi, 
\label{hath}
\ee 
where a constant term has been omitted and 
\be
g_{1D}=g_0\frac{\omega_\perp}{2\pi}=2\omega_\perp a_s,
\ee 
with the scattering length $a_s$ given in the gravitational units (\ref{units}). 

The Hamiltonian (\ref{hath}) is the many-body counterpart of the single-particle Hamiltonian (\ref{h}). Similarly, as in the single-particle case, the energy is not conserved but there exist many-body Floquet eigenstates which evolve periodically with the period of the mirror oscillations, $T$. We will focus on quantum states related to the classical $s:1$ resonant dynamics. For sufficiently strong interaction, the corresponding many-body Floquet eigenstates become extremely  vulnerable to any perturbation because they form macroscopic superpositions (Schr\"odinger cat-like states) and even an infinitesimally weak perturbation, for example, measurement of the position of a single atom, is sufficient to cause the collapse of the Schr\"odinger cat states into one of the states which forms the macroscopic superposition. In the $s:1$ resonant case this means that the discrete time translation symmetry of the Hamiltonian is spontaneously broken into another discrete time translation symmetry \cite{Sacha2015}. In other words, the Floquet eigenstates, which must obey the symmetry of the Hamiltonian, evolve with the period of the mirror oscillations $T$ but the system spontaneously chooses symmetry broken states which evolve with a period $s$ times longer. This phenomenon is dubbed discrete time crystal formation because a quantum many-body system, due to interaction between the particles, spontaneously self-reorganizes its motion and starts evolving with a period different from the period expected from the symmetry of the Hamiltonian \cite{Sacha2015,Khemani16,ElseFTC}. It is in analogy to space crystal formation where atoms, due to mutual interactions, spontaneously form a periodic arrangement in space which breaks the continuous space translation symmetry of the solid state Hamiltonians. In the $2:1$ resonant case ($s=2$), a full many-body analysis was possible and even simulations of the spontaneous symmetry breaking process in the course of the measurements of the position of the particles could be performed \cite{Sacha2015}. Here, we are interested in $s\gg 1$ and the full many-body calculations are not attainable. Therefore, in order to describe experimentally relevant conditions we are going to apply the mean field approach where the spontaneous time translation symmetry breaking is indicated by the loss of   stability of the mean field solutions which evolve with   period $T$ and the emergence of stable solutions evolving with period $sT$. The mean field approach is valid provided that the number of atoms depleted from a Bose-Einstein condensate is much smaller than the total number of particles. It is true in the regime where no spontaneous breaking of time translation symmetry occurs and deep in the discrete time crystal regime where symmetry broken states are Bose-Einstein condensates \cite{Sacha2015,Zin2008}.

The simplest way to switch from the full many-body description to the mean field approach is to exchange the bosonic field operator $\hat\psi(z,t)$ with a classical field $\psi(z,t)$. Then, the Hamiltonian (\ref{hath}) becomes the energy functional of a Bose-Einstein condensate with $\psi(z,t)$ being the condensate wavefunction. We are looking for periodic solutions of the periodically driven system within the mean field approach, and therefore we introduce the energy function related to the Floquet Hamiltonian \cite{Sacha15a}
\bea
E&=&\frac{1}{sT}\int\limits_0^{sT}dt\int\limits_0^\infty dz\;\psi^*\left[\frac{p^2}{2}+z+\lambda z\cos\omega t \right.
\cr &&
\left. +\frac{g_{\rm 1D}N}{2}|\psi|^2-i\partial_t\right]\psi,
\label{efun}
\eea
where $N$ stands for the total number of atoms and we assume that $\int_0^\infty dz|\psi(z,t)|^2=1$. The energy (\ref{efun}) can be considered as the energy of the system per particle averaged over time. Anticipating the emergence of stable solutions with   period $sT$ we have introduced   averaging over a time period $s$ times longer than the driving period $T$. Solving the Gross-Pitaevskii equation (GPE) \cite{Pethick2002} corresponding to (\ref{efun}),
\be
\left[\frac{p^2}{2}+z+\lambda z\cos\omega t+g_{\rm 1D}N|\psi(z,t)|^2-i\partial_t\right]\psi=\mu\psi,
\label{gpfull}
\ee
we can obtain condensate wavefunctions which evolve periodically in time.

In the absence of   particle interactions, the single-particle Floquet eigenstates, related to the classical $s:1$ resonant dynamics, form an $s$-dimensional Hilbert subspace and   are superpositions of $s$ localized wavepackets $w_i(z,t)$ which evolve along the classical $s:1$ resonant orbit. Despite the fact that the wavepackets $w_i(z,t)$ propagate with   period $sT$, the Floquet eigenstates are periodic with  period $T$ because after each period of the mirror oscillations the $w_i(z,t)$'s exchange their positions. The wavepackets $w_i(z,t)$ are   Wannier states corresponding to the first energy band of the effective Hamiltonian (\ref{heff}). Restricting to the $s$-dimentional Hilbert subspace we can expand solutions of the GPE in the Wannier like basis, $\psi(z,t)=\sum_{i=1}^sa_iw_i(z,t)$, and obtain the energy functional (\ref{efun}) within the tight-binding approximation [cf. Eq.~(\ref{1dtight})],
\be
E\approx-\frac{J}{2}\sum\limits_{i=1}^s(a^*_{i+1}a_i+c.c)+\frac12\sum\limits_{i,j=1}^sU_{ij}|a_i|^2|a_j|^2,
\label{mtight}
\ee
where 
\bea
J&=&-\frac{2}{sT}\int\limits_{0}^{sT}dt\int\limits_0^\infty dz\;w_{i+1}^*\left[\frac{p^2}{2}+z+\lambda z\cos\omega t-i\partial_t\right]w_{i}, \cr && \\
U_{ii}&=&\frac{g_{\rm 1D}N}{sT}\int\limits_{0}^{sT}dt\int\limits_0^\infty dz|w_i|^4, \\
U_{ij}&=&\frac{2g_{\rm 1D}N}{sT}\int\limits_{0}^{sT}dt\int\limits_0^\infty dz|w_i|^2|w_j|^2, \quad {\rm for} \quad i\ne j.
\eea
Note that the amplitude $J$ is related to the tunneling of atoms between the wavepackets that are neighbors on the classical trajectory. The tight-binding approximation (\ref{mtight}) is valid provided the interaction energy per particle is much smaller than the energy gap between the first and the second energy bands of the effective Hamiltonian (\ref{heff}) \cite{Dutta2015}.

We have completed all the  necessary theoretical tools to analyze the conditions for experimental realization of various time crystal phenomena.

\section{Discrete time crystals}
\label{spont_DTC}

The formation of space crystals is related to the spontaneous breaking of  continuous space translation symmetry. Hamiltonians of solid state systems do not change if all particles are translated by an arbitrary vector. This symmetry implies that if the systems are prepared in eigenstates, the probability for detection of a single particle must be uniform in space and no crystalline structures are visible. However, exact symmetric ground states of condensed matter systems are strongly vulnerable to a perturbation and even measurement of the position of one particle is sufficient to uncover a crystalline structure in space due to localization of the center of mass of the system. Such a space crystal lives infinitely long if we deal with a macroscopic system because quantum spreading of the center of mass lasts so long that it is not measurable \cite{Sacha2017rev}. 

The Wilczek idea that the same phenomenon can be observed in the time domain if certain quantum many-body systems are prepared in ground states turned out to be impossible \cite{Sacha2017rev}. However, spontaneous breaking of time translation symmetry is possible if one relaxes the requirement and considers excited eigenstates of quantum many-body systems. The so-called discrete time crystals are related to periodically driven systems which spontaneously break discrete time translation symmetries of Hamiltonians and self-reorganize their motion and start evolving with a period different from the period of the driving. The first idea of a discrete time crystal was based on ultra-cold atoms bouncing on an oscillating atom mirror \cite{Sacha2015}. In the present section we are going to analyze the experimental conditions for realization of this system.  

In the original version of the discrete time crystal the $2:1$ resonant dynamics of ultra-cold atoms bouncing on an oscillating mirror was considered \cite{Sacha2015}. From an experimental point of view, in order to reduce atomic losses, it is better to work with a higher resonance. We have chosen the $40:1$ resonance. Discrete time crystal formation in this case means that the system spontaneously self-reorganizes its motion and starts evolving with a period 40 times longer than the period expected from the symmetry of the Hamiltonian. We will first focus on the theoretical considerations and then switch to analysis of the experimental conditions.  

\subsection{Theoretical analysis}
\label{DTCtheory}

\begin{figure} 	            
\includegraphics[width=1.\columnwidth]{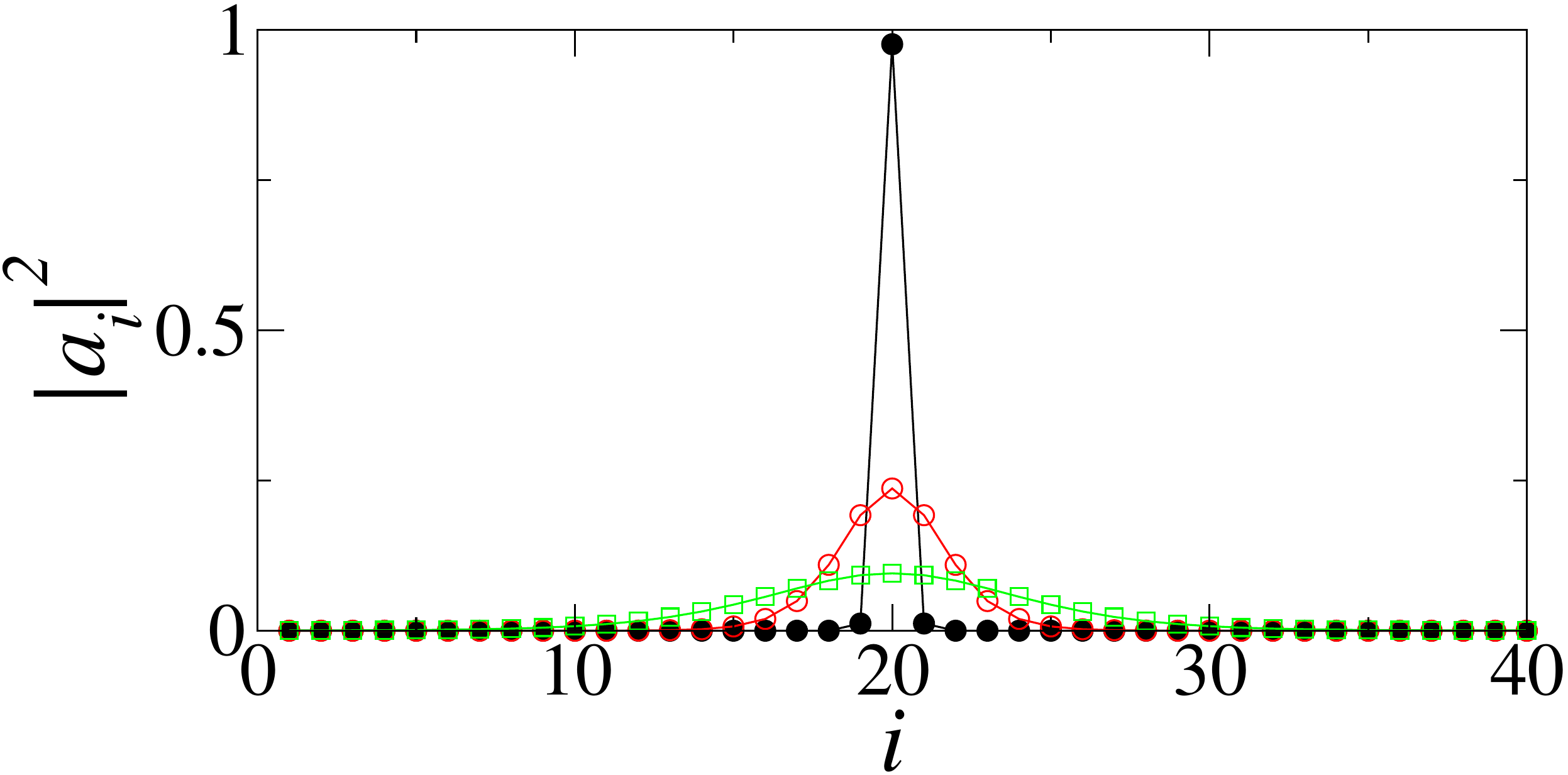}            
\caption{Lowest energy solutions corresponding to the mean field energy functional within the tight-binding approximation (\ref{mtight}) for interaction strengths $g_{\rm 1D}N=-0.12$ (black filled circles), $-0.01$ (red open circles) and $-0.003$ (green open squares). For $g_{\rm 1D}N\gtrsim -1.64\times 10^{-3}$ the symmetry is not broken and the lowest energy solutions are uniform.}
\label{fig2}   
\end{figure} 

\begin{figure*} 	            
\includegraphics[width=0.985\columnwidth]{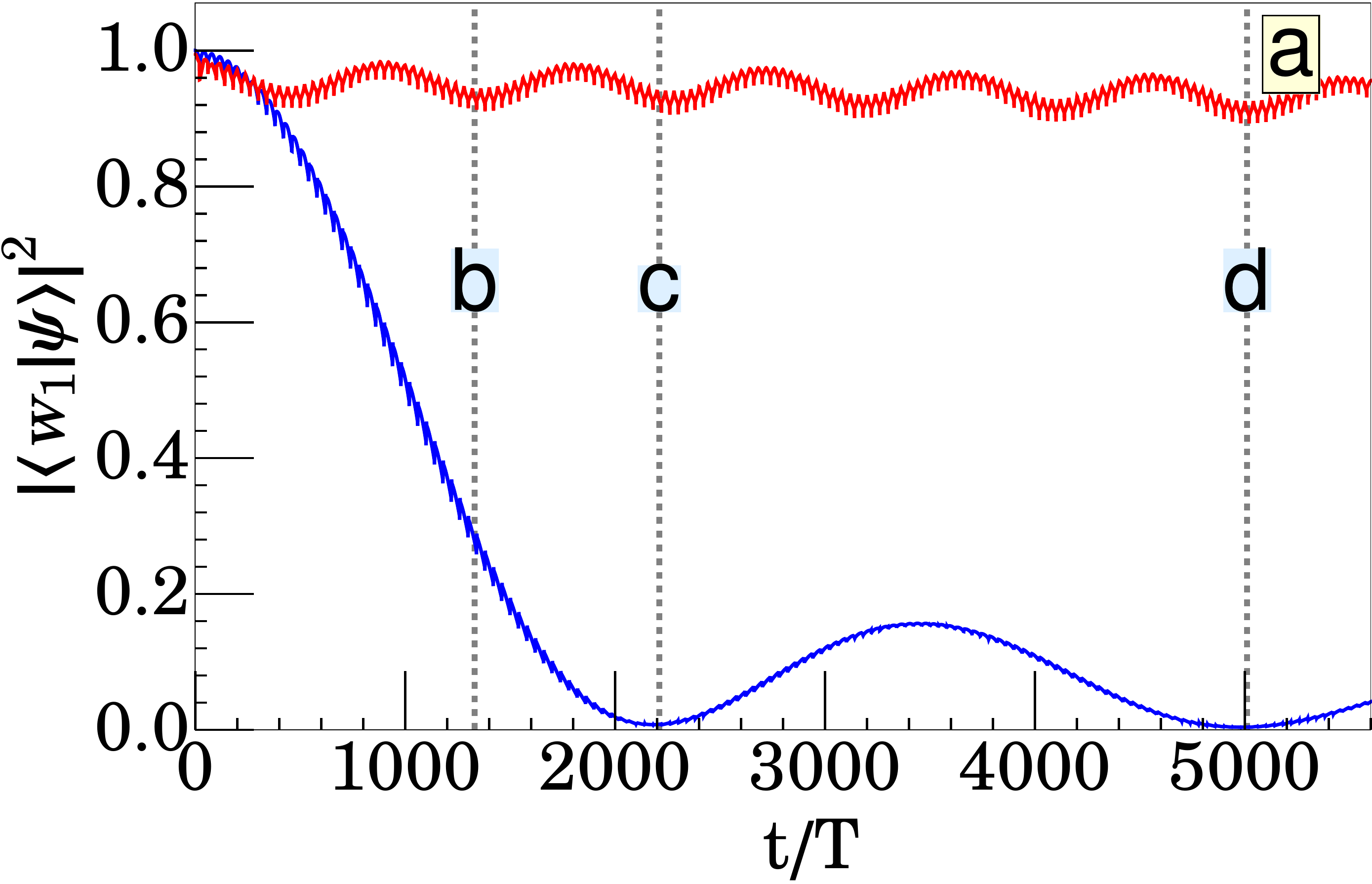}       
\includegraphics[width=1.\columnwidth]{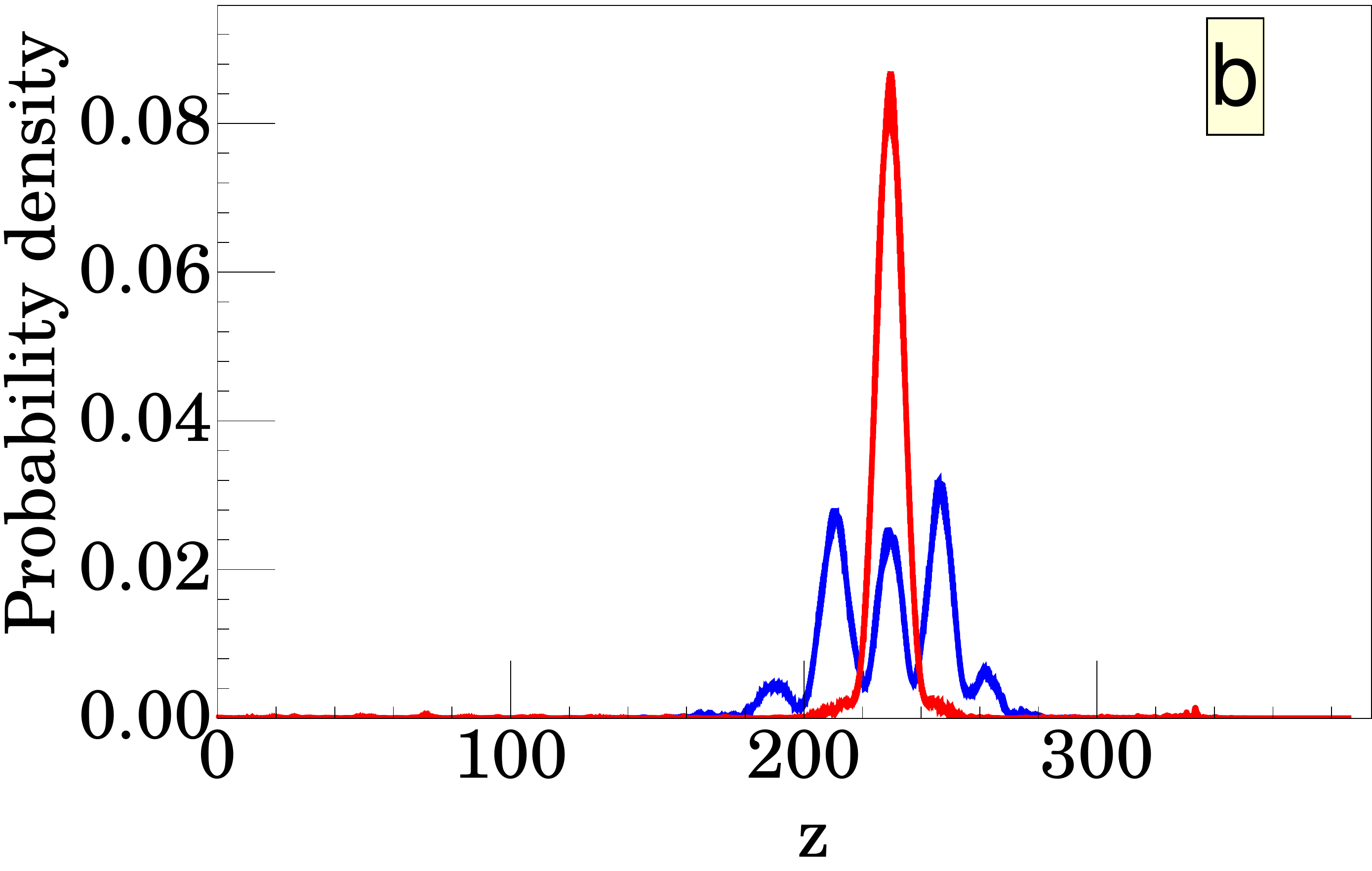}       
\includegraphics[width=1.\columnwidth]{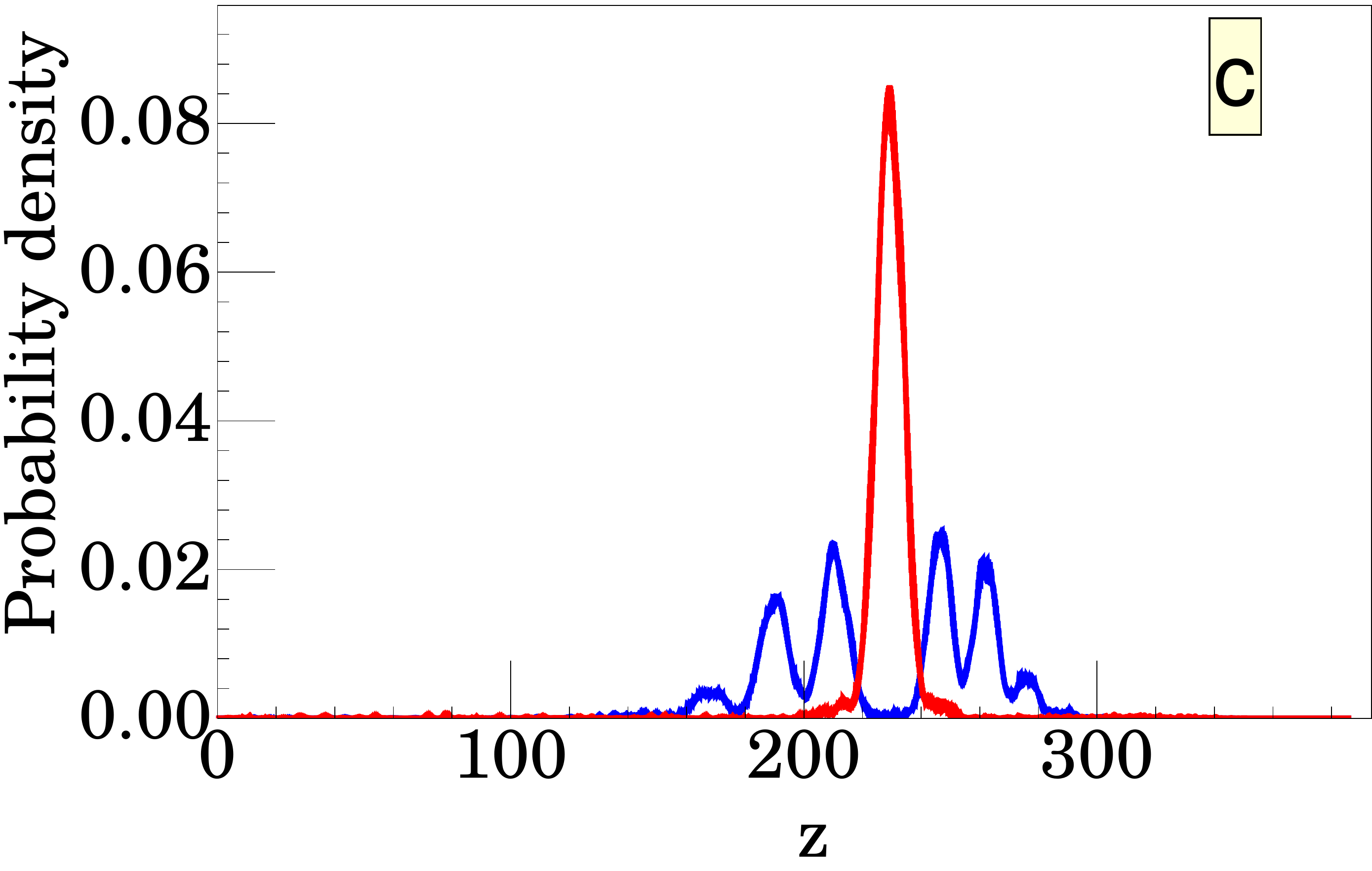}       
\includegraphics[width=1.\columnwidth]{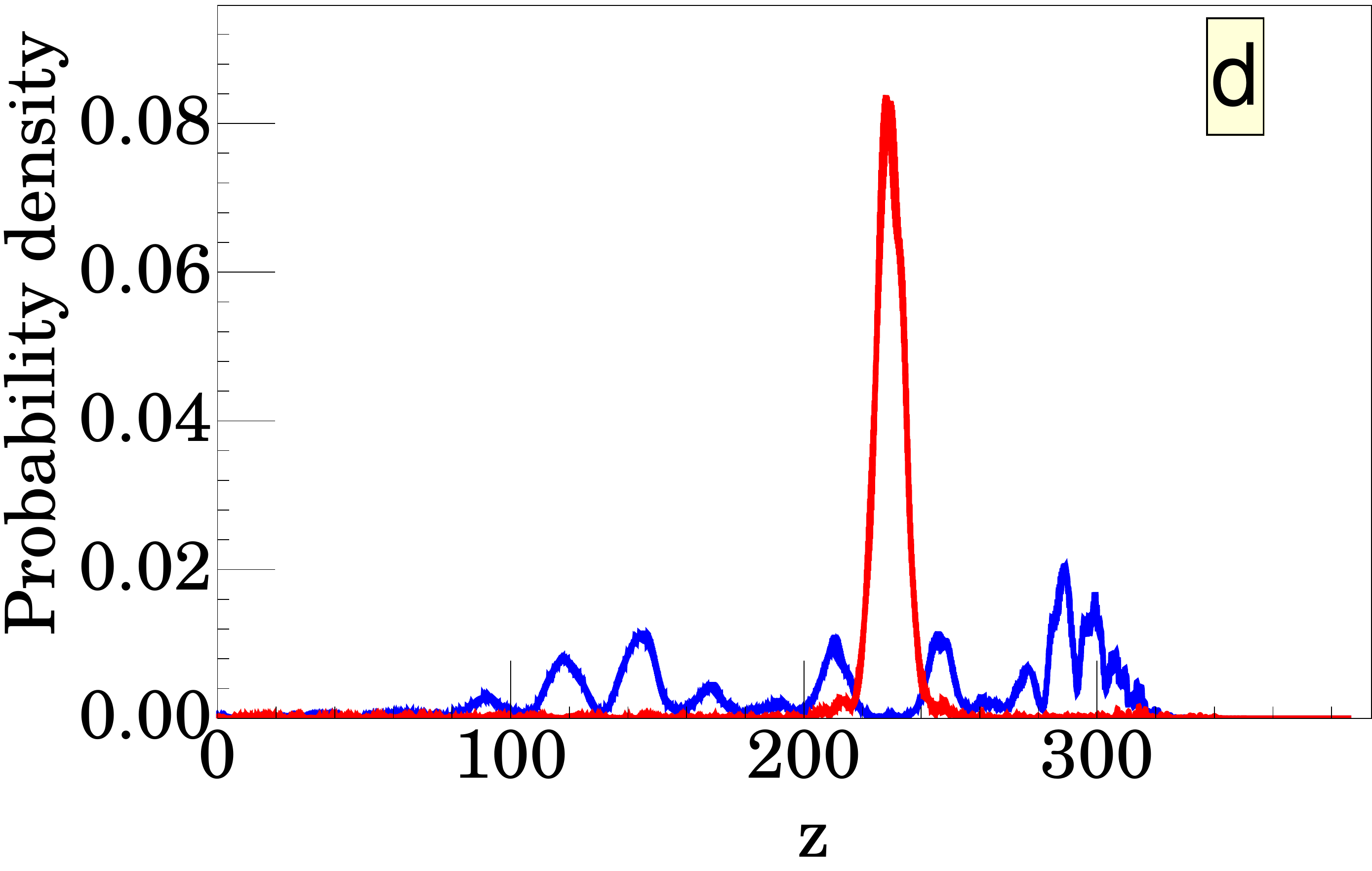}       
\caption{Demonstration of the discrete time crystal formation, see also Fig.~\ref{figoptimal1}. Initially an atomic cloud is prepared in a trap above a periodically oscillating mirror. The parameters of the trap are chosen so that the longitudinal width of the atomic distribution $|\psi(z,0)|^2$ fits the width of the Wannier wavepacket $|w_1(z,0)|^2$ located at the classical turning point. When the longitudinal trapping potential is turned off, atoms fall on the mirror and their time evolution is described by the full GPE (\ref{gpfull}). In the absence of   interactions, atoms tunnel slowly to neighboring Wannier wavepackets which is indicated by a decrease of the overlap between the time evolving wavefunction and the initially chosen Wannier state $|\la w_1(t)|\psi(t)\ra|^2$ [see blue curve in (a)] --- at $t/T=2210$ all atoms tunnel out from the initially chosen Wannier wavepacket and $|\la w_1(t)|\psi(t)\ra|^2=0$. However, when the attractive interactions are turned on ($g_{\rm 1D}N=-0.12$), the system chooses a periodic solution evolving with a period 40 times longer than the period expected from the symmetry of the Hamiltonian and  the discrete time translation symmetry is broken. This situation is presented in (a) by the red curve where the interaction strength $g_{\rm 1D}N=-0.12$. In panels (b)-(d) the densities of interacting (non-interacting) atoms are presented with the red (blue) curves at different moments of time --- we have chosen time moments when the atomic clouds are mid-way between the mirror location ($z=0$) and the classical turning point ($z=329$). Panel (b) corresponds to $t/T=1330$ when in the non-interacting case the initial and neighboring Wannier wavepacktes are equally populated.
 Panel (c) is related to the time moment $t/T=2210$ when all non-interacting atoms tunnel out of the initial Wannier wavepacket while panel (d) corresponds to $t/T=5010$.
During the entire evolution the wavefunction of the interacting system does not decay demonstrating the stability of the discrete time crystal --- see red curves in (b)-(d). 
}
\label{figoptimal}   
\end{figure*} 

A quantum many-body description shows that for sufficiently strong attractive interactions between the ultra-cold atoms, low-lying energy states in the Hilbert subspace related to the classical $2:1$ resonant dynamics possess Schr\"odinger cat-like structures and the measurement of the position of a particle in the system prepared in one of them breaks discrete time translation symmetry and a discrete time crystal forms \cite{Sacha2015}. This phenomenon can also be described by means of the mean field theory which indicates a loss of the stability of the condensate wavefunction propagating with the period of the mirror oscillation $T$ and the emergence of new stable solutions evolving with the period $2T$ \cite{Sacha2015}. We apply the latter approach to the $40:1$ resonance case.

The simplest way to identify the critical strength of the interactions between atoms that leads to the breaking of  discrete time translation symmetry is to analyze the lowest energy solution of the energy functional in the tight-binding approximation (\ref{mtight}). From the classical analysis presented in Fig.~\ref{fig1} we know that for $s=40$ the exact phase portrait is reproduced by the prediction of the effective Hamiltonian (\ref{heff}) for $\lambda=0.2$. Having determined $\lambda$ we need to choose the frequency of the mirror oscillation, $\omega$, or equivalently the resonant action $I_{40}\equiv I_{s=40}$. The latter determines an effective Planck constant $I_{40}^{-1}$ which has to be smaller than the area of one of the elliptical resonant islands visible in Fig.~\ref{fig1}. From an experimental point of view we need to ensure that the tunneling time of atoms between the elliptical islands, $2.4/J$, is much shorter than the lifetime of a Bose-Einstein condensate in the laboratory. On the other hand, we cannot afford $J$ to be too large because then the energy gap between the first and the second energy bands of the effective single particle Hamiltonian (\ref{heff}) is very small and when we turn on the particle interactions the simple picture of the time crystal formation, described by the single band tight-binding model (\ref{mtight}), is lost. A compromise is to  choose $\omega$  such that the energy gap is at least of the order of $10J$. Simple diagonalization of the single-particle effective Hamiltonian (\ref{heff}) shows that for $\omega=4.9$ (i.e., $I_{40}=1790$) the energy gap equals $9.5J$, where $J=8.6\times10^{-4}$.

Having chosen the parameters of the single-particle problem we can now analyze what strength of the attractive interactions is needed to break the symmetry and which value is suitable for an experiment. For $g_{\rm 1D}=0$, the energy (\ref{mtight}) is minimized by the uniform solution, $a_i=1/\sqrt{40}$, which corresponds to a wavefunction $\psi(z,t)$ that evolves with the period $T$. For $g_{\rm 1D}N\lesssim-1.64\times 10^{-3}$ the uniform solution becomes unstable and new degenerate stable solutions are born which evolve with the period $40T$. For $|g_{\rm 1D}N|$ greater than the critical value but close to it, the lowest energy solution is slightly non-uniform and it could be difficult to prepare and detect it experimentally. Therefore, it is much better to increase $|g_{1D}N|$ so that the symmetry broken solutions are localized in single sites of the tight-binding problem (\ref{mtight}). In Fig.~\ref{fig2} we present the solutions for different interaction strengths. The value of about $g_{1D}N=-0.12$ results in the symmetry broken solutions being nearly entirely localized in single sites which correspond to the interaction energy per particle $|U_{ii}|/2=3.8J$. The latter is 2.5 times smaller than the energy gap between the first and second energy bands which guarantees the validity of the tight-binding approximation. 

In  three dimensional space a Bose gas can collapse if the attractive contact interactions are sufficiently strong \cite{Castin_LesHouches,Saito2001,Saito2001a,Donley2001,Altin2011}. In order to prevent such a bosenova effect a transverse trap has to be present and the number of atoms cannot be too large. We have already assumed that in the transverse directions the condensate wavefunction $\phi_0(x,y)$ corresponds to the ground state of the harmonic trap of   frequency $\omega_\perp$. We will see that a discrete time crystal is represented by a wavepacket $\psi(z,t)$ evolving along a classical trajectory whose probability distribution reveals the smallest standard deviation  at the classical turning point, which we denote by $\sigma$. The system will be stable against  collapse if the resulting interaction energy per particle at the turning point,  $g_0N\int d^3r|\phi_0\psi|^4$, is smaller than the excitation energy along the transverse directions, $\omega_\perp$, that leads to the following condition \cite{Castin_LesHouches}
\be
\sigma \gtrsim |a_s|N.
\ee

The present subsection has allowed us to determine the optimal parameters for the formation of a discrete time crystal on the basis of the $40:1$ resonant dynamics. In the next subsection we analyze the experimental conditions and perform time-dependent numerical simulations of the realization of the discrete time crystal by integrating the full GPE (\ref{gpfull}), i.e., without referring to the tight-binding approximation.

\begin{figure} 	            
\includegraphics[width=1.\columnwidth]{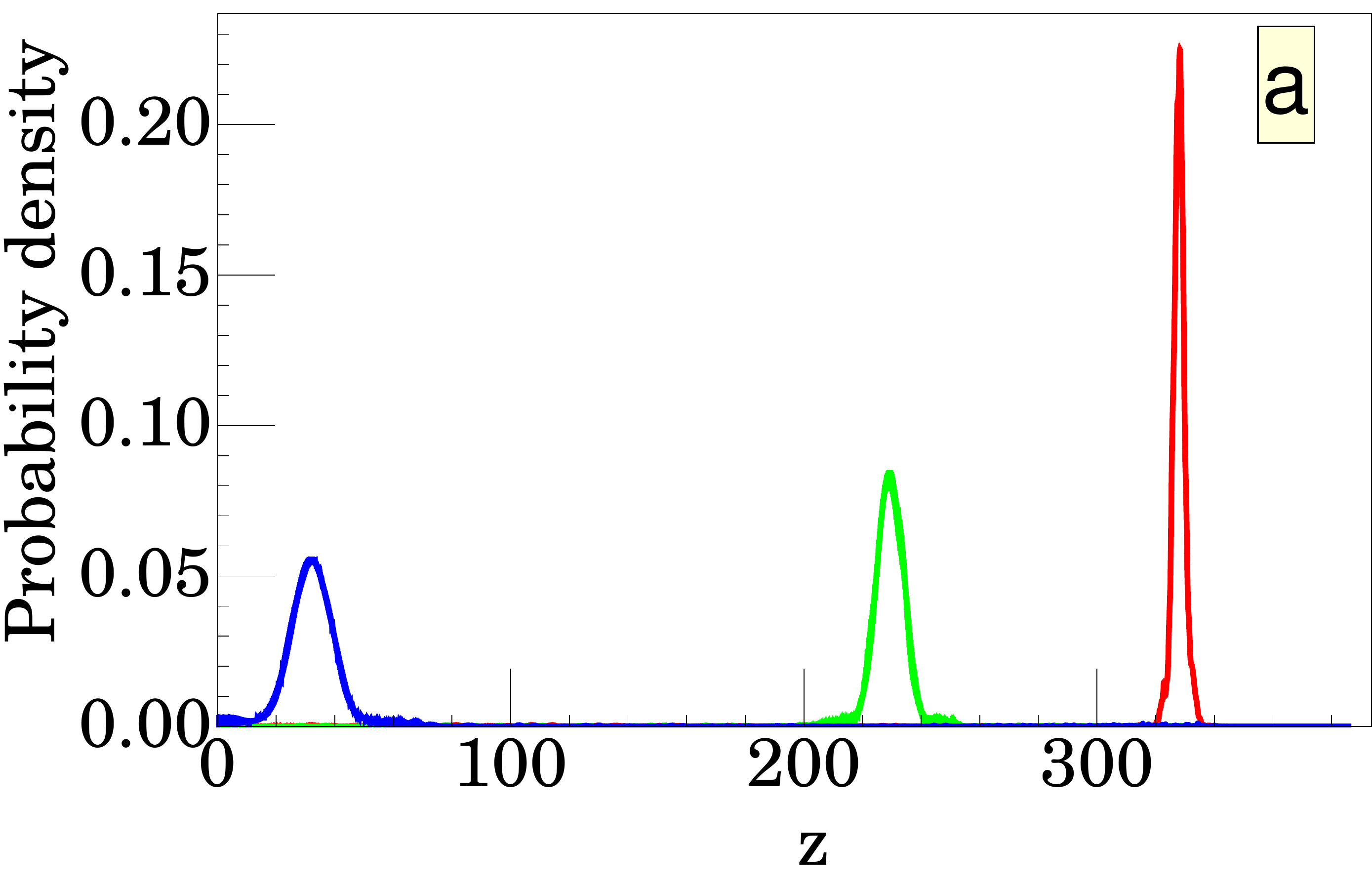}       
\includegraphics[width=1.\columnwidth]{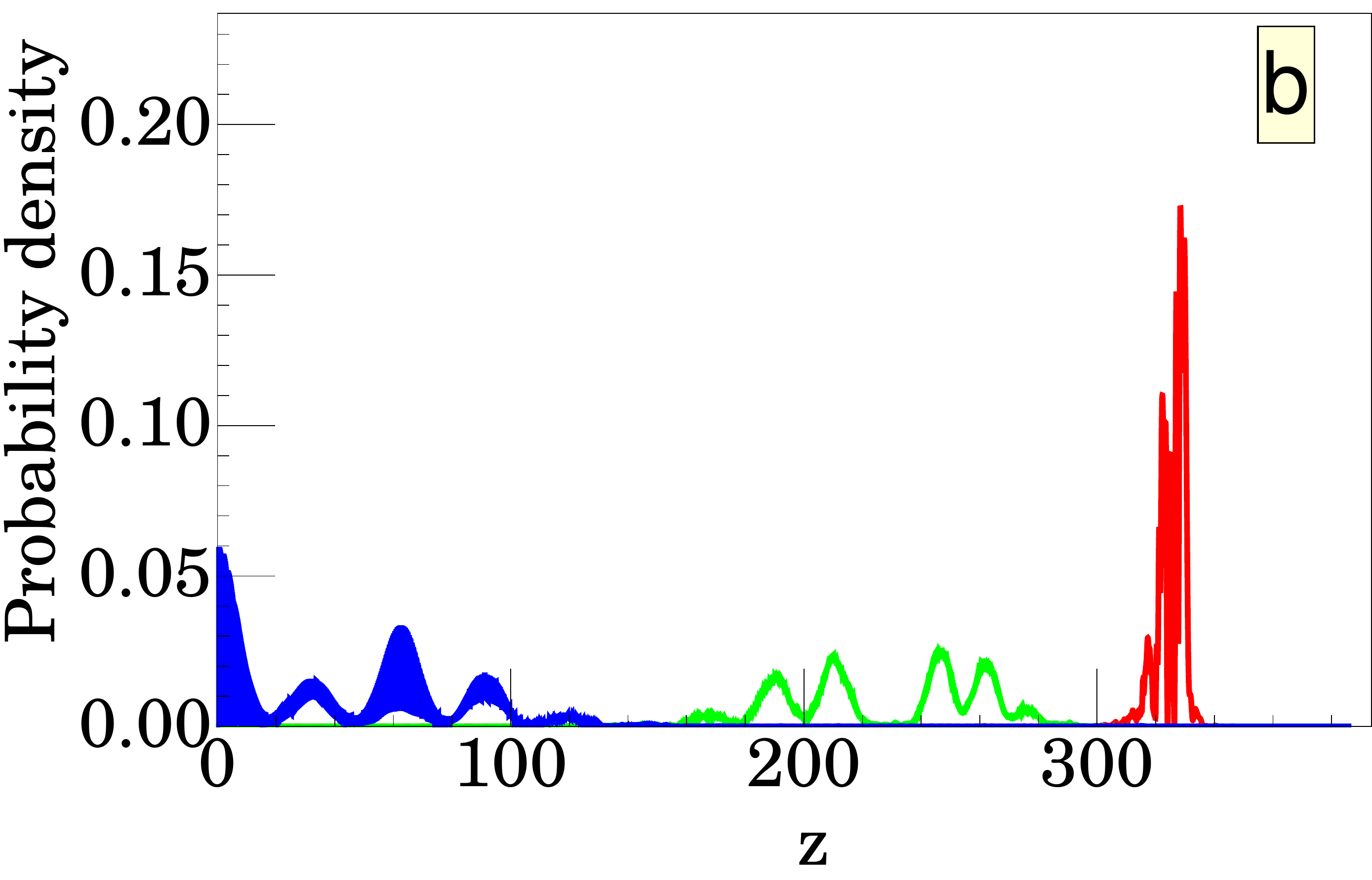}       
\caption{Experimental signatures of the formation of the discrete time crystal. A periodically driven quantum many-body system, due to the interactions between particles, does not follow the driving force but spontaneously chooses motion with a period 40 times longer than the period of the driving. This is illustrated in panel (a) where we plot the densities of the atomic cloud around $t/T=2210$, i.e, at $t/T=2200$ (red), 2210 (green) and 2220 (blue) for $g_{\rm 1D}N=-0.12$. The time crystal behavior is in contrast to the evolution of the system without interactions. Panel (b) shows that without   interactions the evolution of the initially localized wavepacket is not stable because atoms have tunneled to other localized wavepackets moving along the classical resonant orbit. The filled-in blue regions between z=0 and z=100 are interference fringes related to the reflection of the wavepackets from the mirror. In both panels the same initial state, described in Fig.~\ref{figoptimal}, has been chosen.
}
\label{figoptimal1}   
\end{figure} 

\begin{figure} 	            
\includegraphics[width=1.\columnwidth]{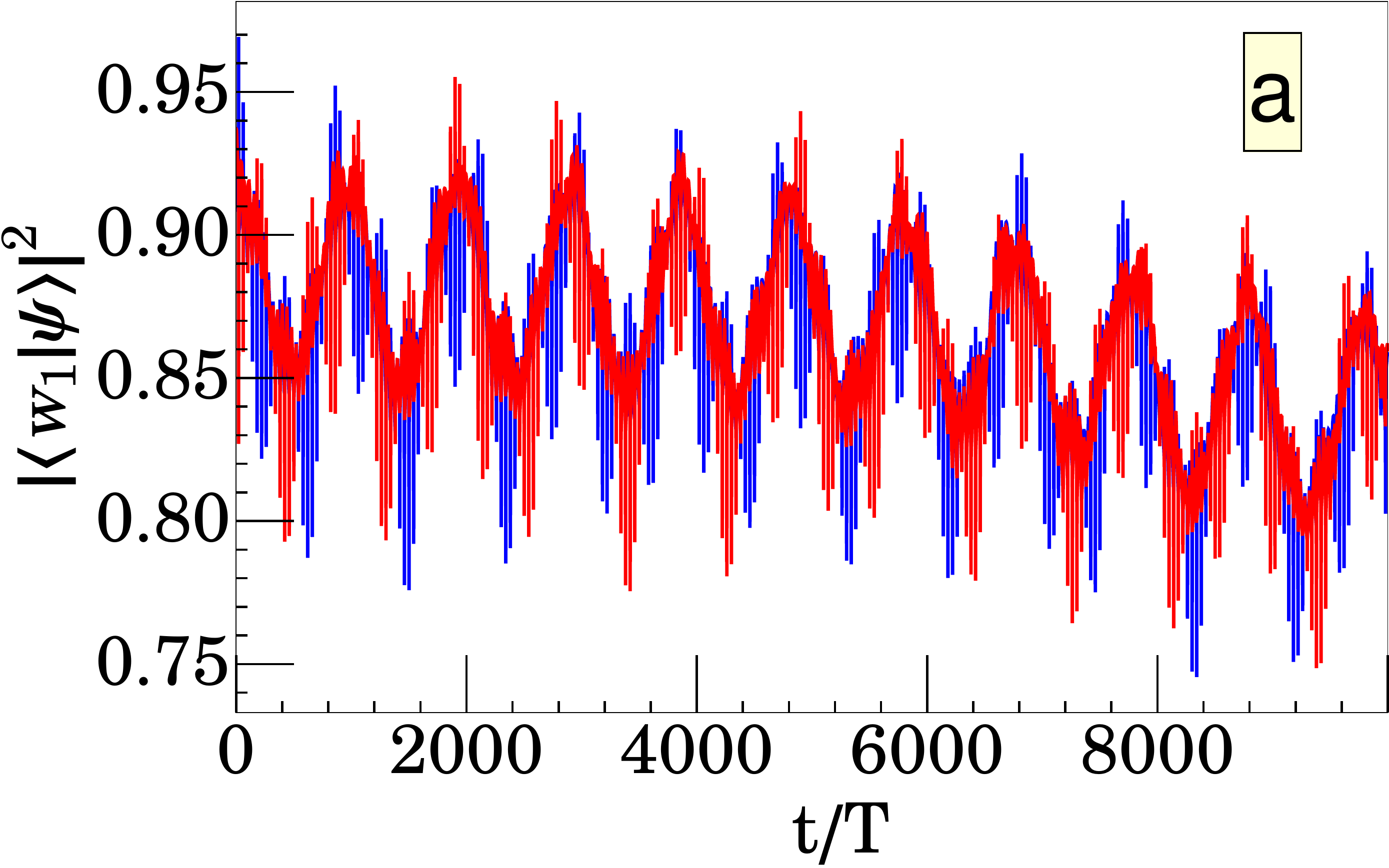}       
\includegraphics[width=1.\columnwidth]{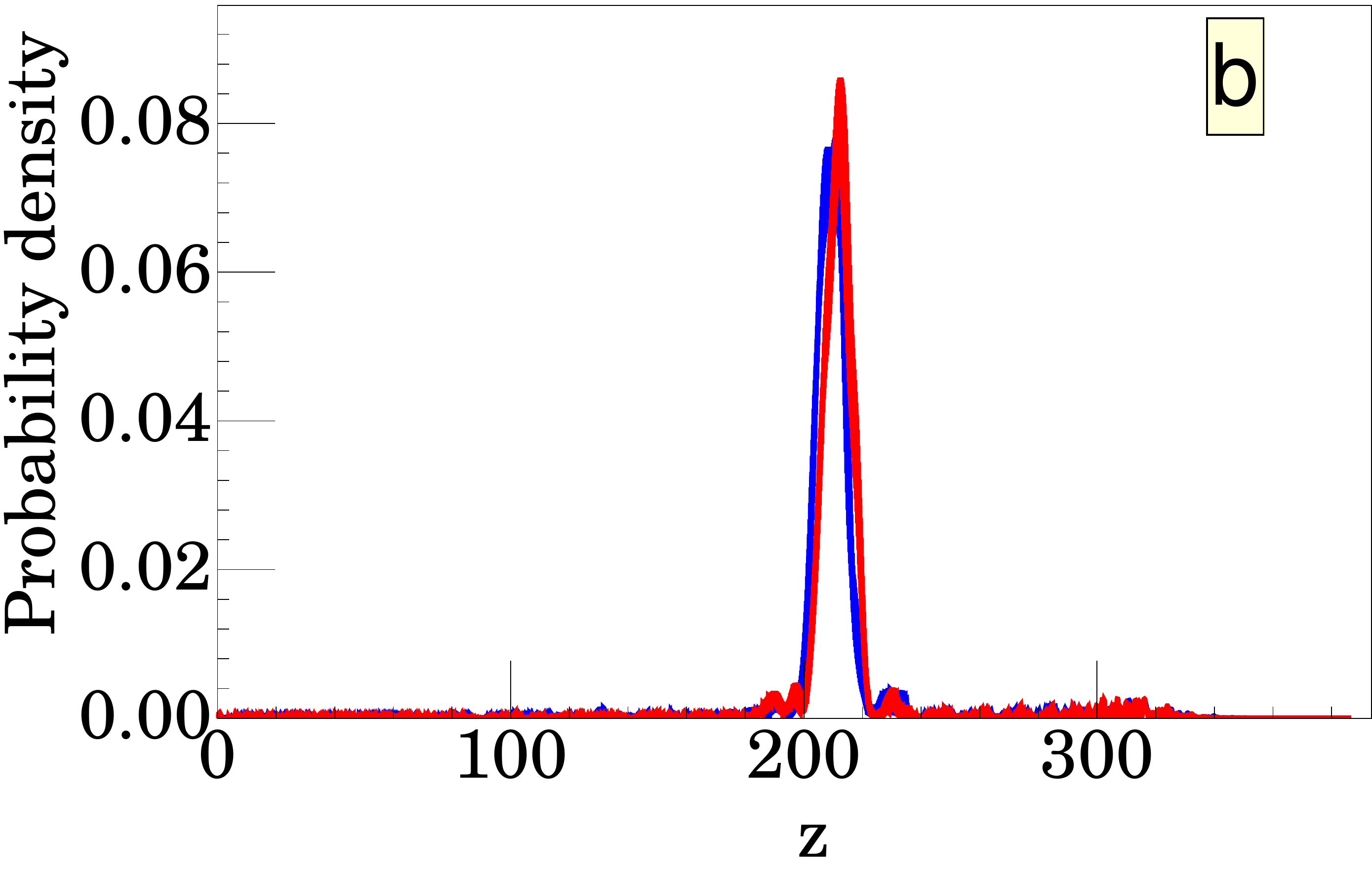}       
\caption{Similar data to that presented in Fig.~\ref{figoptimal} in the interacting case ($g_{\rm 1D}N=-0.12$); however, the initial atomic distribution $|\psi(z,0)|^2$ is displaced along the longitudinal direction with respect to the classical turning point by $+0.5\sigma$ (red curves) and $-0.5\sigma$ (blue curves), where $\sigma$ is the standard deviation of the probability density $|\psi(z,0)|^2$. Panel (a) shows the corresponding $|\la w_1(t)|\psi(t)\ra|^2$ while panel (b) presents $|\psi(z,t)|^2$ at $t/T=9990$.
}
\label{figdev}   
\end{figure} 

\subsection{Experimental conditions}

A suitable atomic system for performing a discrete time crystal experiment based on ultra-cold atoms bouncing on an oscillating mirror is the 
$^{85}\mbox{Rb}$ $|F=2,m_F=-2\ra$ state, which has a broad Feshbach resonance that allows precise tuning of the s-wave scattering length and hence the interparticle interaction \cite{Claussen2003}. Using gravitational units for $^{85}\mbox{Rb}$, $l_0=0.385 \, \mu \mbox{m}$ and $t_0 = 0.198 \, \mbox{ms}$, we obtain the following values for the parameters determined in the previous subsection: mirror oscillation frequency  $\omega/(2\pi t_0) = 3.94 \, \mbox{kHz}$ and amplitude $\gamma l_0 = 3.2\, \mbox{nm}$, distance of the classical turning point from the mirror $h l_0 = 127 \,\mu \mbox{m}$, and standard deviation of the atomic density along the longitudinal direction at the classical turning point $\sigma l_0 = 0.77 \, \mu \mbox{m}$. The maximal allowed number of atoms is $N \leq \sigma/|a_s|$, which for $a_sl_0 \approx -0.1 \, \mbox{nm}$ ($-2a_0$) yields $N \leq 8000$. The desired interaction strength, $g_{\rm 1D} N = 2 \omega_{\perp} a_sN = -0.12$, can be achieved by a proper choice of $\omega_{\perp}$, $a_s$ and $N$; e.g., $a_sl_0 = -0.1\, \mbox{nm}$, $\omega_{\perp}/(2\pi t_0) = 37\,\mbox{Hz}$ and $N = 5000$. These parameters are considered to be realistic for an experiment.

In the original proposal of a discrete time crystal based on ultra-cold atoms bouncing on an oscillating mirror \cite{Sacha2015}, we considered the simple case of a $2:1$ resonance in which the initial state was a single Floquet state consisting of a superposition of $s=2$ wavepackets. In order to keep the required number of bounces to a reasonable number to reduce possible atom losses, it is better to work with a higher resonance, such as the $40:1$ resonance. However, it is difficult to prepare a single Floquet state for the $40:1$ resonance because it is a superposition of 40 localized wavepackets moving with different velocities and with certain mutual phase relations. On the other hand, it is relatively easy to prepare a single localized wavepacket that moves periodically along the $40:1$ resonant orbit and to monitor the evolution of the wavepacket with and without the attractive interactions.   

The experiment starts with the ground state of a BEC of $^{85}\mbox{Rb}$  atoms in a three-dimensional harmonic optical dipole trap whose center is located at the classical turning point. The frequencies of the (pancake shape) trap are  $\omega_{\perp}/(2\pi t_0) = 37\,\mbox{Hz}$ and  $\omega_{\parallel}/(2\pi t_0) = 100\,\mbox{Hz}$ along the transverse and longitudinal directions, respectively. The value of the longitudinal trap frequency ensures that the standard deviation of the ground-state distribution along the longitudinal direction at the classical turning point is $\sigma l_0 = 0.77 \, \mu \mbox{m}$. 

Once the initial state of the system is prepared, we turn off the trapping potential along the longitudinal direction and the atom cloud starts falling on the oscillating atom mirror under the influence of gravity and the tightly confining one-dimensional red-detuned optical waveguide. The frequency of the mirror oscillations is adjusted to the frequency of the classical $40:1$ resonant orbit which depends on the initial distance of the cloud from the mirror. The initial phase of the mirror oscillation corresponds to $t = 0$ in Eq.~\eqref{mirror} and hence the mirror needs to be in the lowest position with zero velocity. Both adjustments can be carried out by tuning the driving of the mirror.

We now analyze the above optimal parameters for the case of the $40:1$ resonance. In Figs.~\ref{figoptimal}-\ref{figoptimal1} we present the evolution of the atomic cloud, both in the presence of the desired attractive interactions ($g_{\rm 1D} N = -0.12$) and in the absence of the interactions. Without the interactions, atoms prepared in the localized wavepacket tunnel slowly to neighboring localized wavepackets which move on the $40:1$ resonant classical trajectory, as indicated by the overlap between the time evolving wavefunction and the initially chosen Wannier state $\left|\la w_1(t) | \psi(t) \ra\right|^2$ (Fig.~\ref{figoptimal}(a), blue line). After $t/T =2210$ mirror periods (i.e., after 55 bounces of the atom cloud on the mirror) no atoms remain in the initial wavepacket (Figs.~\ref{figoptimal}(c), blue line and Fig.~\ref{figoptimal1}(b), green line) because all the atoms have tunneled to neighboring wavepackets. In the presence of sufficiently strong attractive interactions the initial Gaussian wavepacket does not decay --- it evolves freely along the classical trajectory with a period 40 times longer than the mirror oscillation period for long times without tunneling losses to other wavepackets [Fig.~\ref{figoptimal}, red lines and Fig.~\ref{figoptimal1}(a)].

The $^{85}\mbox{Rb}$ BEC is prepared by setting the s-wave scattering length to  $a_sl_0\approx 300 a_0$ by means of the broad Feshbach resonance at the magnetic field $155.041(18)\,\mbox{G}$ \cite{Claussen2003} and sympathetically cooling with a $^{87}\mbox{Rb}$ BEC in an optical dipole trap \cite{Kuhn2014}. Once the BEC is reached, the evaporative cooling can be continued which allows one to reduce the thermal cloud and to obtain the desired number of atoms in the trap. The atom mirror is a blue-detuned repulsive light sheet created by focusing 532 nm light, e.g., from a 10 W frequency-doubled Nd:YVO$_4$ laser, with a cylindrical lens \cite{Bongs1999}. The bouncing atoms are confined in a single transverse mode of the tightly confining red-detuned optical waveguide. Such an atom mirror can be modulated with the required frequency of $\sim$4 kHz and amplitude of $\sim$3 nm by vibrating a beam-guiding optical mirror with a piezo crystal, by modulating the beam with an AOM or by modulating the optical potential. A similar light-sheet atom mirror has previously been used to bounce a BEC of $^{87}\mbox{Rb}$ atoms dropped from heights of $\sim 100 \, \mu\mbox{m}$ by Bongs et al. \cite{Bongs1999}, who demonstrated the coherent evolution of the bouncing BEC. For the $40:1$ resonance and the above parameters,
about 55 bounces, or about 0.6~s, are required without significant loss of atoms or loss of phase coherence to allow time for the atoms to tunnel to neighboring wavepackets in the absence of interactions.

The evolution of the atom density of the bouncing $^{85}\mbox{Rb}$ atoms is monitored when the s-wave scattering length is adiabatically changed from $\approx 300 a_0$ to: 
\begin{enumerate}
 \item[(i)] zero, to turn off the interactions to allow the atoms to start to tunnel to other wavepackets in times of order $1/J$. To control the scattering length to $\pm0.1a_0$ requires the magnetic field to be stable to about $\pm$2 mG; 
 \item[(ii)] $a_sl_0=-2a_0$, corresponding to an attractive interaction $g_{\rm 1D} N=-0.12$ for $N=5000$. This is sufficiently strong to break the time-translation symmetry to form a discrete time crystal which evolves freely with period $40T$ for long times without tunneling to other wavepackets. The stability of the discrete time crystal can be tested by introducing controlled fluctuations of the mirror amplitude.
\end{enumerate}

We now consider the influence of possible experimental imperfections:

\begin{enumerate}
 \item Precise control of the total number of atoms $N$, the s-wave scattering length $a_s$ and the transverse confinement frequency $\omega_{\perp}$ is not important because all of these parameters influence the value of $g_{\rm 1D}$ only. For example, when $g_{\rm 1D}$ is changed by 10\%, the results presented in Fig.~\ref{figoptimal}(a) do not change.
 \item Precise control of the frequency $\omega_{\parallel}$ of the initial longitudinal trapping potential is also not essential. When the width of the initial atomic distribution along the longitudinal direction is changed by 10\% with respect to the optimal value, the results shown in Fig.~\ref{figoptimal}(a) do not change.
 \item Precise location of the initial atomic distribution at the desired classical turning point is not important because deviations can be corrected by an appropriate adjustment of the frequency of the mirror oscillations. The important factor is the stability of the location of the atom distribution in different realizations of the experiment. In Fig.~\ref{figdev} we show the results when the initial distribution is displaced with respect to the optimal position by 0.5 times its standard deviation, i.e., by $0.5\sigma l_0 = 0.38\,\mu\mbox{m}$. There is a drop of the squared overlap by only a few percent. Thus, in different realizations of the experiment the allowed deviations of the location of the initial distribution with respect to the mirror position are of the order of 0.5$\sigma$. 
\end{enumerate}

\section{Dynamical quantum phase transition}
\label{DQPT}

In equilibrium statistical physics  phase transitions are indicated by a non-analytical behavior of the macroscopic quantities as a function of a control parameter \cite{Sachdev2011,Dziarmaga2010}. It turns out that a similar non-analytical behavior can also be observed as a function of time in the non-equilibrium dynamics of  many-body systems which is induced by a quantum quench, i.e., a sudden change of a control parameter across a critical value \cite{Heyl2013,Heyl2018rev}. Such a dynamical quantum phase transition is indicated by a non-analyticity of the return probability of a system to the initial state at a certain critical moment of time if the thermodynamic limit is considered. This phenomenon can be interpreted as the partial loss of information on the system evolution --- when we observe a system from the point of view of the initial state there is a breakdown of the short-time expansion at a critical moment of time. While dynamical quantum phase transitions have been analyzed mostly in time independent systems, recently a similar behavior has been predicted in a periodically driven system which reveals discrete time crystal formation \cite{Kosior2017}. Here, we analyze the experimental signatures of a dynamical quantum phase transition for ultra-cold atoms bouncing on a periodically oscillating mirror after a quantum quench from the time crystal phase to a weakly interacting regime.

\begin{figure} 	            
\includegraphics[width=0.9\columnwidth]{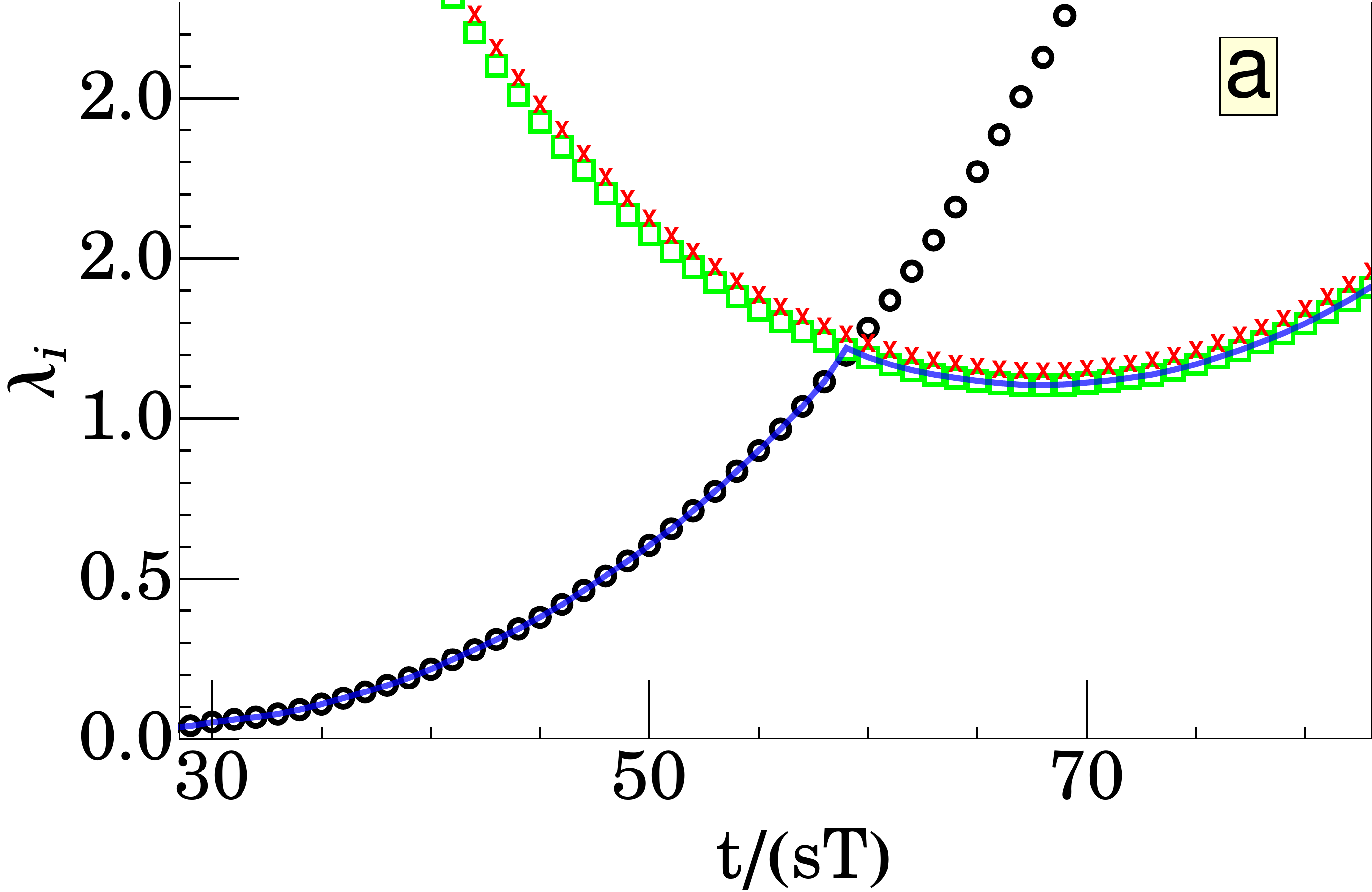}       
\includegraphics[width=0.91\columnwidth]{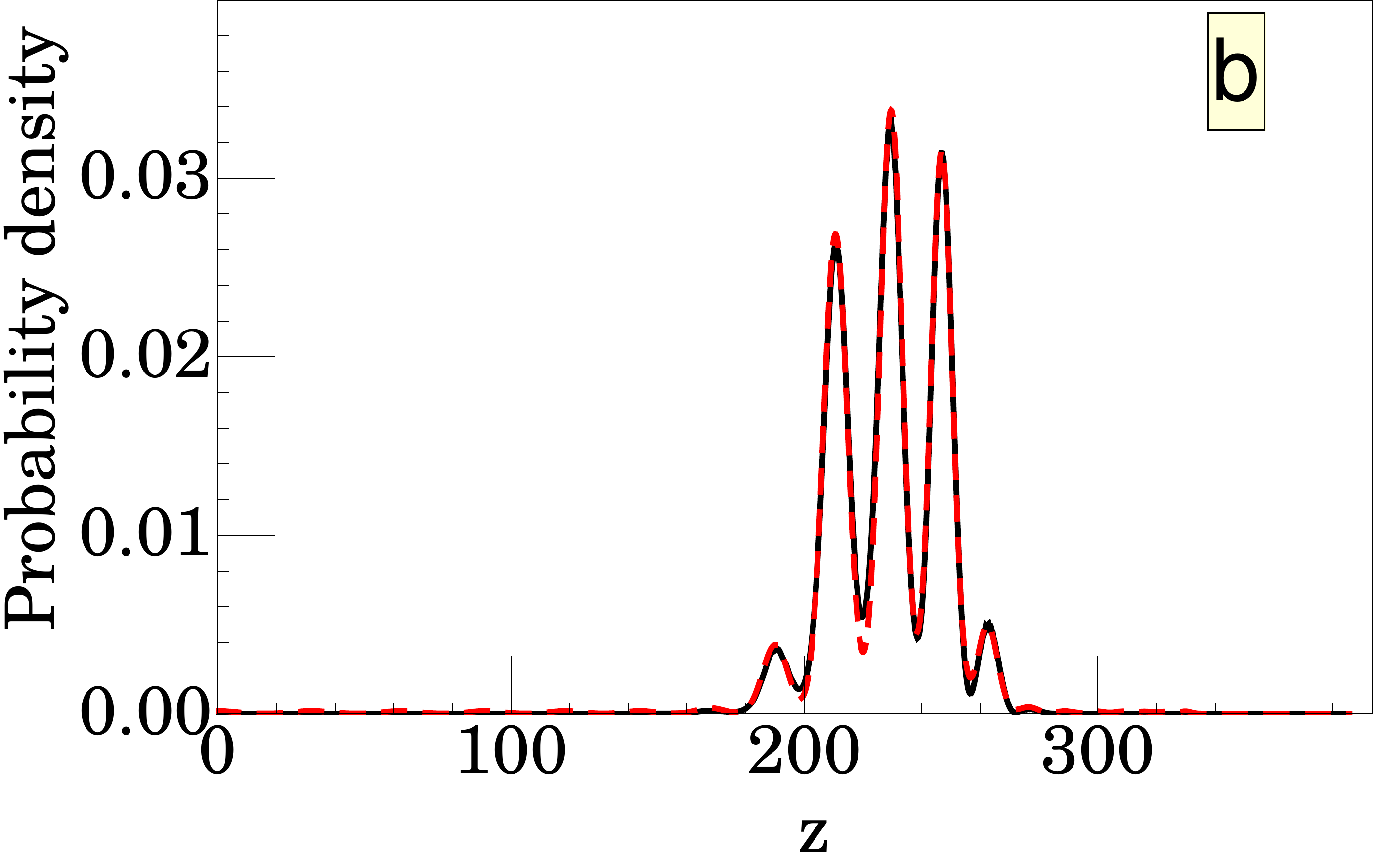}       
\caption{Dynamical quantum phase transition in the discrete time crystal. Initially, ultra-cold atoms are prepared in a harmonic trap above an oscillating mirror and at $t=0$ the longitudinal trapping potential is turned off and atoms fall on the mirror and form a discrete time crystal, i.e., they break the time translation symmetry and evolve with a period $s=40$ longer than $T$. At time $t/(sT)=28$ the interactions between atoms are turned off and the system evolves according to a different Hamiltonian and the time crystal decays. Panel (a) shows the evolution of the smallest three rates $\lambda_i(t)$, sampled with the period $sT=40T$, obtained by fitting a sum of Gaussian distributions to the atomic densities --- each Gaussian represents a Wannier-like wavepacket $|w_i(z,t)|^2$. Black symbols correspond to the overlap of $\tilde\psi(z,t)$ with the initial Wannier-like wavepacket $w_1(z,t)$, while red crosses and green squares correspond to the overlap with the two neighboring Wannier functions. Blue solid line is related to the minimal rate $\lambda_{\rm min}(t)$. The crosses and squares are not identical because the state prepared initially in the trap is not a perfect Wannier function $w_1(z,t)$.  An example of the result of the fitting is presented in panel (b), where we have chosen $t$ close to the critical time $t_c/(sT)=59$, i.e., at the moment of time when initially the smallest rate approaches two other rates corresponding to the projection of $\tilde\psi(z,t)$ on two neighboring Wannier-like wavepackets. In panel (b) the solid black line corresponds to $|\tilde\psi(z,t)|^2$ and the dashed red line to the fitted sum of the Gaussian distributions.
}
\label{figrates}   
\end{figure} 

A quantum many-body description of ultra-cold bosons bouncing resonantly on an oscillating mirror can be reduced to a Bose-Hubbard Hamiltonian. That is, when we restrict to the Hilbert subspace spanned by Fock states $|n_1,\dots,n_s\ra$, where $n_i$ denotes the number of bosons occupying a Wannier-like wavepacket $w_i(z,t)$ evolving along the $s:1$ resonant orbit, we end up with a many-body Bose-Hubbard Hamiltonian similar to Eq.~(\ref{mtight}) but with the complex numbers $a_i$ replaced by the standard bosonic annihilation operators $\hat a_i$. If the attractive interactions between bosons are sufficiently strong, the low-lying energy eigenstates of the Bose-Hubbard Hamiltonian are Schr\"odinger cat-like states. For example, the ground state $|\Psi\ra$ can be approximated by a macroscopic superposition $(|N,0,\dots,0\ra+|0,N,0,\dots,0\ra+\dots+|0,\dots,0,N\ra)/\sqrt{s}$. For $s=2$ it was shown that starting with $|\Psi\ra$ and performing a quench to a weakly interacting regime, 
the so-called Loschmidt echo ${\cal L}(t)=|\la\Psi(t)|\tilde\Psi(t)\ra|^2$, where $|\Psi(t)\ra$ and $|\tilde\Psi(t)\ra$ evolve according to the Hamiltonians before and after the quench, respectively, reveals a cusp at a critical moment of time $t_c$ \cite{Kosior2017}. In an experiment it is extremely difficult to prepare a Schr\"odinger cat state because the system spontaneously chooses a symmetry broken state and a discrete time crystal forms, which has been analyzed in Sec.~\ref{spont_DTC}. However, the signature of the dynamical quantum phase transition can still be observed experimentally even if we prepare initially a discrete time crystal state, e.g., $|\Psi\ra=|\Phi_1\ra=|N,0,\dots,0\ra$, where all bosons occupy a single Wannier-like wavepacket $w_1(z,t)$, and after the quench we perform measurements of the return probability of the system to the lowest energy manifold. Indeed, in the limit when $N\rightarrow\infty$, the return probability to the degenerate ground-state manifold of the Bose-Hubbard Hamiltonian reproduces the Loschmidt echo \cite{Heyl2018rev,Kosior2017}, i.e.,
\be
{\cal L}(t)=\sum\limits_{i=1}^s p_i(t)=\sum\limits_{i=1}^s e^{-N\lambda_i(t)},
\label{returnp}
\ee
where $p_i(t)=|\la\Phi_i(t)|\tilde\Psi(t)\ra|^2$ is the probability of finding the evolving state $|\tilde\Psi(t)\ra$ in a time crystal state $|\Phi_i(t)\ra$ where $N$ bosons occupy a Wannier-like wavepacket $w_i(z,t)$. In (\ref{returnp}) we have introduced the so-called rates $\lambda_i(t)$ which are intensive quantities and read 
\be
\lambda_i(t)=-\frac{1}{N}\ln p_i(t).
\label{rates}
\ee
In the thermodynamic limit, the Loschmidt echo is determined by the smallest rate at a given moment of time $t$, 
\be
{\cal L}(t)\propto  e^{-N\lambda_{\rm min}(t)},
\ee
where $\lambda_{\rm min}(t)={\rm min}\{\lambda_1(t),\dots,\lambda_s(t)\}$ \cite{Heyl2018rev}. A non-analytical behavior of ${\cal L}(t)$ corresponds to a time moment when initially the smallest rate $\lambda_i(t)$ has increased so much that it becomes greater than another rate $\lambda_j(t)$. Then, the minimal rate $\lambda_{\rm min}(t)$ reveals a cusp which corresponds to a cusp of the Loschmidt echo ${\cal L}(t)$, see Fig.~\ref{figrates}(a).
Experimentally, it is very difficult to measure directly the Loschmidt echo because ${\cal L}(t)$ drops exponentially with $N$. However, the rates themselves can be measured and the dynamical quantum phase transition can be observed in the laboratory \cite{Jurcevic2017,Flaschner2018}.  

The experiment demonstrating a dynamical quantum phase transition in a discrete time crystal can be performed with the parameters determined in Sec.~\ref{spont_DTC}. Initially, for the interaction strength $g_{\rm 1D}N=-0.12$, one prepares a discrete time crystal state $|\Psi(t)\ra\approx|\Phi_1(t)\ra$ which evolves with a period $s=40$ times longer than the mirror oscillation period $T$, see Figs.~\ref{figoptimal}-\ref{figoptimal1}. At a certain moment of time $t$, the interactions are instantly turned off ($g_{\rm 1D}=0$) and the system starts evolving according to a new Hamiltonian. We are interested in the rates (\ref{rates}) where $p_i(t)$ are the probabilities of finding the evolving state $|\tilde\Psi(t)\ra$ in the different time crystal states $|\Phi_i(t)\ra$. Due to the fact that the initial state and the time crystal states $|\Phi_i(t)\ra$ are Bose-Einstein condensates where macroscopic numbers of atoms occupy Wannier-like wavepackets $w_i(z,t)$, the entire experiment can be described by the GPE (\ref{gpfull}). The projections of the solution $\tilde\psi(z,t)$ of the GPE on the Wannier-like wavepackets allows us to calculate the probabilities $p_i(t)=|\la w_i(t)|\tilde\psi(t)\ra|^{2N}$ and consequently the rates $\lambda_i(t)=-\ln |\la w_i(t)|\tilde\psi(t)\ra|^2$. In an experiment, the rates can be obtained by measuring atomic densities. Indeed, by determining the fractions of the total number of atoms which form a given localized wavepacket we obtain estimates for $|\la w_i(t)|\tilde\psi(t)\ra|^2$. This is illustrated in Fig.~\ref{figrates} where we present the rates $\lambda_i(t)$ obtained by fitting a sum of Gaussian distributions to the atomic densities. The crossing point of the smallest rates corresponds to the critical time $t_c$ when the return probability of the system (\ref{returnp}) reveals a cusp indicating the dynamical quantum phase transition.

\section{Anderson localization in time}
\label{AL_in_time}

In this section we still analyze ultra-cold atoms bouncing on an oscillating mirror but we do not consider spontaneous breaking of time translation symmetry and formation of discrete time crystals. 

In Sec.~\ref{system} we showed that atoms bouncing resonantly on the mirror behave like electrons in a space crystal and such a crystalline behavior is observed in the time domain. In this section we show that when we introduce disorder in the driving, Anderson localization phenomena known in condensed matter physics can be realized in time.

Anderson localization is a well known phenomenon which takes place in configuration space and relies on an exponential localization of eigenstates of a particle \cite{Anderson1958,MuellerDelande:Houches:2009}. When a time-independent spatially periodic potential is contaminated by a spatially disordered contribution, extended Bloch waves turn into exponentially localized eigenstates due to destructive interference between different multiple scattering paths. The localization of eigenstates  is accompanied by the inhibition of transport in a disordered system. Anderson localization has also been studied in disordered systems with fast periodic time modulations \cite{Kosior2015,Major2016,Major2017}. Actually, the presence of a spatially periodic potential is not necessary. Indeed, even without a crystalline structure in space, the presence of a disordered potential characterized by a finite correlation length results in the localization of a particle in configuration space. Anderson localization can also be observed in momentum space and it is related to the quantum suppression of classical diffusion of classically chaotic systems \cite{Fishman:LocDynAnders:PRL82,Casati:IncommFreqsQKR:PRL89,Lemarie:Anderson3D:PRA09}. Yet another kind of Anderson localization has recently been proposed: localization in the time domain due to the presence of disorder in time \cite{Sacha15a,sacha16,Giergiel2017,delande17}. 

In this section we show that non-interacting ultra-cold atoms bouncing on an atom mirror can reveal Anderson localization in time if the mirror performs random motion.
Experimental signatures of the localization are related to an exponential localization, around a certain moment of time, of the probability for the detection of atoms at a fixed position on a resonant periodic orbit \cite{Sacha15a}.

\begin{figure} 	            
\includegraphics[width=0.94\columnwidth]{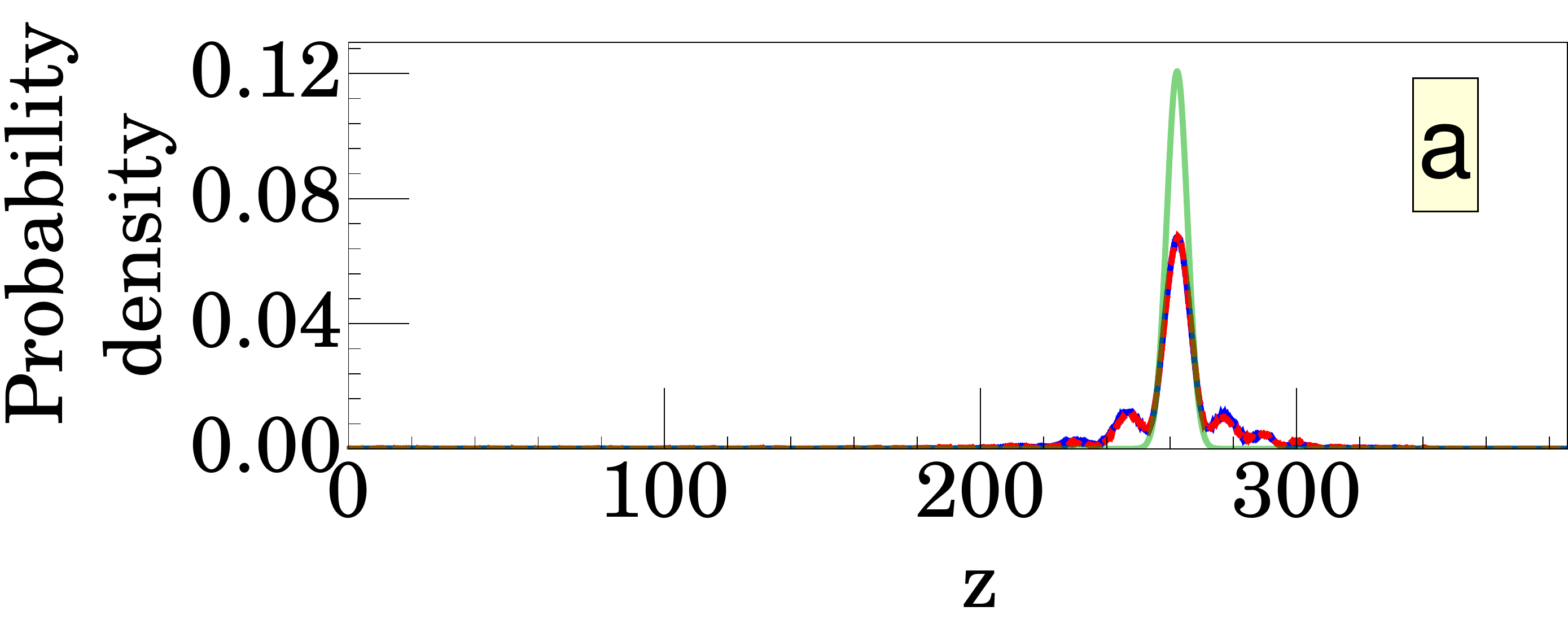}       
\includegraphics[width=1.\columnwidth]{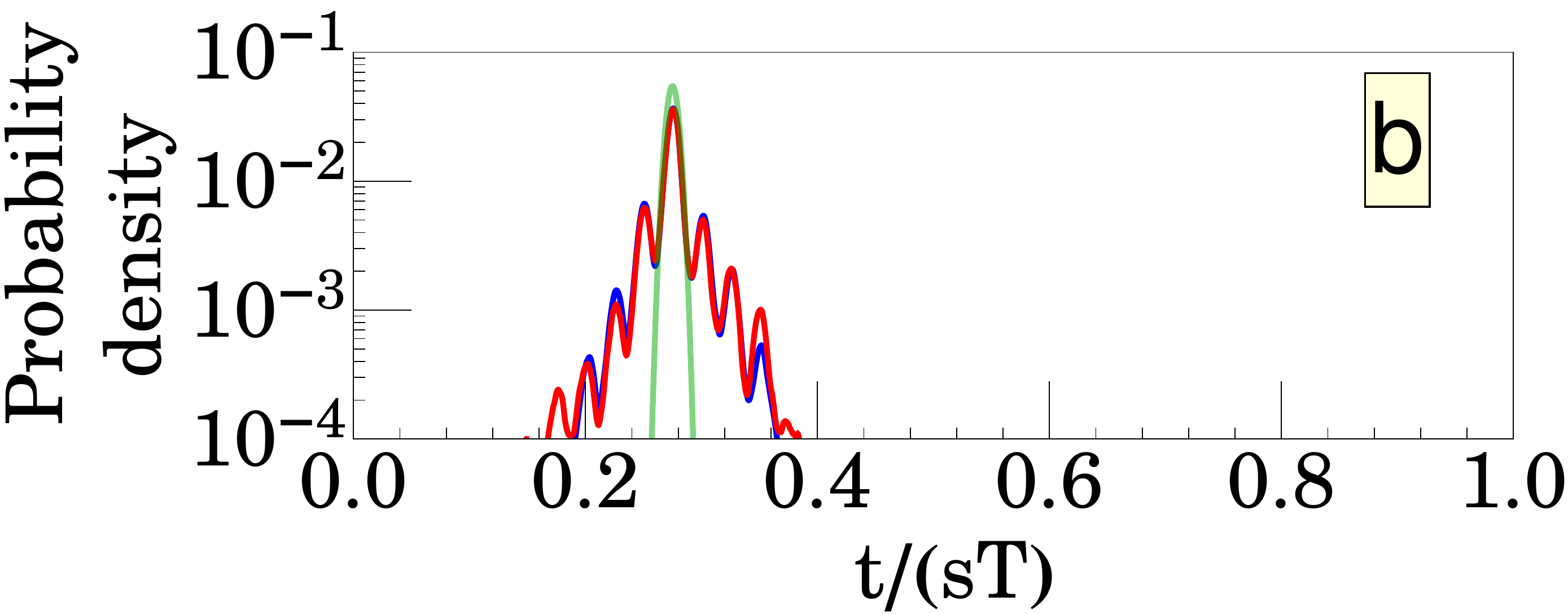}       
\caption{Anderson localization in the time domain in the presence of crystalline structure in time. Panel (a) shows the initial atomic density (green curve) and the densities averaged over 50 realizations of the random motion of the mirror, see Eq.~(\ref{frandom1}), at long evolution times, i.e., $t=2210T$ (blue curve) and $t=5010T$ (red curve). Panel (b) presents the Anderson localization in the time domain, i.e., the average probability densities for the detection of atoms at $z=1$ versus time. Green, blue and red curves correspond to the results presented in panel (a), i.e., to the probability of the detection of atoms around $t=0$, $t=2210T$ and $t=5010T$, respectively. Once the Anderson localization freezes the spreading of the wavepacket, the probability density for the detection of atoms at $z=1$ does not change its shape and it is repeated with period $sT$ due to the periodic boundary conditions in time --- the presented results correspond to $s=40$.}
\label{al1}   
\end{figure} 

We begin with an atom bouncing on a harmonically oscillating mirror where we assume that the $s:1$ resonance condition is fulfilled. For $s\gg 1$, the eigenstates of the effective single-particle Hamiltonian (\ref{heff}) are Bloch waves which in the laboratory frame appear as trains of localized Wannier-like wavepackets, $w_i(z,t),$ moving periodically along a classical resonant orbit. 
 When we restrict to the first energy band of the effective Hamiltonian (\ref{heff}), the energy of an atom can be described by the tight-binding model (\ref{1dtight}).

Let us assume that the motion of the mirror is not perfectly harmonic because we turn on a fluctuating perturbation. That is, the mirror motion is no longer described by Eq.~(\ref{mirror}) but by 
\bea
f(t)&=&f(t+sT) \cr
&=&-\gamma\cos\omega t-\frac{V_d}{\sqrt{2s}}\sum\limits_{\substack{k=-s\\k\ne 0}}^s e^{i(k\omega t/s+\varphi_k)},
\label{frandom1}
\eea
where $T=2\pi/\omega$ and the phases $\varphi_k=-\varphi_{-k}$ are chosen randomly from the uniform distribution in the interval $[0,2\pi]$. The first part of $f(t)$ is responsible for the crystalline structure described by (\ref{heff}) while the other part is the fluctuating perturbation whose strength is characterized by $V_d$. For the parameters chosen in Sec.~\ref{spont_DTC}, the first part corresponds to the main driving with frequency $\omega/(2\pi t_0)=3.94$~Hz and the perturbation part consists of its $s=40$ sub-harmonics. The presence of the perturbation results in additional terms in the effective Hamiltonian (\ref{heff}) which currently reads
\be
H_{\rm eff}\approx \frac{P^2}{2m_{\rm eff}}+V_0\cos(s\Theta)+\frac{V_d}{\sqrt{2s}}\sum\limits_{\substack{k=-s\\k\ne 0}}^s e^{i(k\Theta+\varphi_k)},
\label{heffdis}
\ee
with $m_{\rm eff}=-\pi^2s^4/\omega^4$ and $V_0=\gamma(-1)^{s+1}$. If we restrict to the first energy band we obtain the tight-binding model
\be
E\approx-\frac{J}{2}\sum\limits_{i=1}^s\left(a_{i+1}^*a_i+c.c.\right)+\sum\limits_{i=1}^s \epsilon_i|a_i|^2,
\label{1dtightdis} 
\ee
where $\epsilon_i=(sT)^{-1}\int_0^{sT}dt\int_0^\infty dzzf''(t)|w_i|^2$. Employing the central limit theorem it can be shown that the randomness of the phases $\varphi_k$ implies that the $\epsilon_i$ are random numbers corresponding to a normal distribution with standard deviation $V_d$.
The tight-binding model (\ref{1dtightdis}) is actually a one-dimensional Anderson model and for $s\rightarrow\infty$ Anderson localization takes place regardless of how small is $V_d$. In the case when $s=40$, which we are interested in, localization can be observed provided eigenstates localize on a number of neighboring Wannier-like wavepackets smaller than 40. For the parameters described in Sec.~\ref{spont_DTC}, the amplitude of the harmonic oscillations of the mirror is $\gamma l_0=3.2$~nm, which results in $J=8.6\times 10^{-4}$, and in order to analyze the Anderson localization we choose $V_d=5.1J$, which corresponds to $V_dl_0=1.7$~nm. 

The experimental realization of Anderson localization in time can start with precisely the same initial state as considered in Sec.~\ref{spont_DTC} and \ref{DQPT}, i.e., with ultra-cold atoms trapped above the mirror. However, in the present case the interactions between atoms have to be turned off and the mirror performs the motion described by (\ref{frandom1}). The initial state is a superposition of the exponentially localized eigenstates described by (\ref{heffdis}) and in the course of time evolution it does not spread over the entire resonant periodic orbit but tends to a localized periodically evolving probability distribution \cite{Sanchez2007,Billy2008,Roati2008}. We have performed numerical integration of the full Schr\"odinger equation of an atom bouncing on a harmonically oscillating mirror and in the presence of random fluctuations, see (\ref{frandom1}). In Fig.~\ref{al1}(a) we show the density of atoms in configuration space averaged over 50 different realizations of the disordered motion of the mirror at $t=2210T$ and $t=5010T$. The results indicate that after sufficiently long evolution the atomic density freezes its shape --- it essentially stops spreading already at time $t$ of the order of $1000T$. This behavior is in contrast to the evolution without the mirror fluctuations ($V_d=0$) presented in Fig.~\ref{figoptimal} and Fig.~\ref{figoptimal1}(b) where tunneling of atoms does not stop and atoms tend to spread along the entire classical resonant orbit. 

Anderson localization in the time domain is illustrated in Fig.~\ref{al1}(b) where the average probability density for the detection of atoms at $z=1$ is shown as a function of time. Once the atomic distribution stops spreading, the localization of the probability density around a certain moment of time is observed and it is repeated with the period $sT$ due to the periodic boundary conditions in time.

The experimental setup considered in this subsection can also be used for realization of many-body localization where interactions between particles and strong disorder result in the absence of thermalization of a system, vanishing of DC transport and logarithmic growth of the entanglement entropy \cite{Basko06,Oganesyan07,Znidaric08,Huse14,Rahul15,kozarzewski2016,mierzejewski2016,Sierant2018}. This phenomenon can be observed in the time domain when together with the presence of the fluctuating atom mirror, repulsive interactions between atoms are turned on \cite{Mierzejewski2017}.

\section{Summary and conclusions}
\label{concl}

In the present paper we have considered various aspects of crystalline behavior in time of ultra-cold atoms bouncing on an oscillating atom mirror and performed a detailed analysis of the realistic experimental conditions needed for their realization. Experiments with cold and ultra-cold atoms bouncing on a mirror have already been carried out \cite{Steane95,Roach1995,Sidorov1996,Westbrook1998,Lau1999,Bongs1999,Sidorov2002,Fiutowski2013,Kawalec2014}. Such a system turns out to be a promising platform for realization of time crystals. 

We began with considerations of the formation of discrete time crystals \cite{Sacha2015,Khemani16,ElseFTC,Yao2017,Lazarides2017,Russomanno2017,Zeng2017,Nakatsugawa2017,Ho2017,Huang2017,Gong2017,Wang2017,Mizuta2018}. Discrete time crystals have already been demonstrated in the laboratory with the help of spin systems \cite{Zhang2017,Choi2017,Nayak2017,Pal2018,Rovny2018,Rovny2018a,Autti2018}. Atoms bouncing on a mirror are able to reveal a dramatic breaking of discrete time translation symmetry of the system Hamiltonian where the symmetry broken states evolve with periods tens of times longer than the driving period. Such discrete time crystals cannot be realized in spin systems unless the spin quantum numbers are very large. We identified experimentally realistic conditions and performed numerical simulations of the formation of a discrete time crystal. We also performed an analysis of the influence of experimental imperfections on the realization of the phenomenon.

Quantum many-body systems which form discrete time crystals can reveal a dynamical quantum phase transition after suddenly turning off the interactions between particles \cite{Kosior2017}. We demonstrated that ultra-cold atoms bouncing on an oscillating mirror are also a suitable system for realization of such a non-analytical behavior in time. Indeed, we showed that measurements of atomic densities after a quantum quench (i.e., after a sudden turn-off of the interactions) allow one to obtain the return probability of the system to the initial time crystal manifold which reveals non-analytical behavior around a critical moment of time.

Atoms resonantly bouncing on a mirror are also a promising system for realization of various condensed matter phenomena in the time domain \cite{Sacha2017rev,Giergiel2018}. We focused on Anderson localization and showed that when the mirror fluctuates in time, the randomness in the driving of the atoms results in Anderson localization of the system in the time domain \cite{Sacha15a,sacha16,Giergiel2017,delande17}. The latter phenomenon corresponds to an exponential localization in time of the probability for the detection of atoms at a fixed position in configuration space. Analysis of the considered system showed that the observation of Anderson localization in time is attainable experimentally.

\vspace{0.01cm}

\section*{Acknowledgement}

We thank Jia Wang, Andrei Sidorov, Nick Robins and Carlos Kuhn for fruitful discussions. Support of the National Science Centre, Poland via Projects No.~2016/20/W/ST4/00314 (K.G.) 2016/21/B/ST2/01086 (A.K.) and under QuantERA programme No.~2017/25/Z/ST2/03027 (K.S.) is acknowledged. 


\begin{thebibliography}{89}
\expandafter\ifx\csname natexlab\endcsname\relax\def\natexlab#1{#1}\fi
\expandafter\ifx\csname bibnamefont\endcsname\relax
  \def\bibnamefont#1{#1}\fi
\expandafter\ifx\csname bibfnamefont\endcsname\relax
  \def\bibfnamefont#1{#1}\fi
\expandafter\ifx\csname citenamefont\endcsname\relax
  \def\citenamefont#1{#1}\fi
\expandafter\ifx\csname url\endcsname\relax
  \def\url#1{\texttt{#1}}\fi
\expandafter\ifx\csname urlprefix\endcsname\relax\def\urlprefix{URL }\fi
\providecommand{\bibinfo}[2]{#2}
\providecommand{\eprint}[2][]{\url{#2}}

\bibitem[{\citenamefont{{Sacha} and {Zakrzewski}}(2018)}]{Sacha2017rev}
\bibinfo{author}{\bibfnamefont{K.}~\bibnamefont{{Sacha}}} \bibnamefont{and}
  \bibinfo{author}{\bibfnamefont{J.}~\bibnamefont{{Zakrzewski}}},
  \bibinfo{journal}{Rep. Prog. Phys.} \textbf{\bibinfo{volume}{81}},
  \bibinfo{pages}{016401} (\bibinfo{year}{2018}),
  \urlprefix\url{https://doi.org/10.1088/1361-6633/aa8b38}.

\bibitem[{\citenamefont{Wilczek}(2012)}]{Wilczek2012}
\bibinfo{author}{\bibfnamefont{F.}~\bibnamefont{Wilczek}},
  \bibinfo{journal}{Phys. Rev. Lett.} \textbf{\bibinfo{volume}{109}},
  \bibinfo{pages}{160401} (\bibinfo{year}{2012}),
  \urlprefix\url{http://link.aps.org/doi/10.1103/PhysRevLett.109.160401}.

\bibitem[{\citenamefont{Bruno}(2013)}]{Bruno2013b}
\bibinfo{author}{\bibfnamefont{P.}~\bibnamefont{Bruno}},
  \bibinfo{journal}{Phys. Rev. Lett.} \textbf{\bibinfo{volume}{111}},
  \bibinfo{pages}{070402} (\bibinfo{year}{2013}),
  \urlprefix\url{http://link.aps.org/doi/10.1103/PhysRevLett.111.070402}.

\bibitem[{\citenamefont{Watanabe and Oshikawa}(2015)}]{Watanabe2015}
\bibinfo{author}{\bibfnamefont{H.}~\bibnamefont{Watanabe}} \bibnamefont{and}
  \bibinfo{author}{\bibfnamefont{M.}~\bibnamefont{Oshikawa}},
  \bibinfo{journal}{Phys. Rev. Lett.} \textbf{\bibinfo{volume}{114}},
  \bibinfo{pages}{251603} (\bibinfo{year}{2015}),
  \urlprefix\url{http://link.aps.org/doi/10.1103/PhysRevLett.114.251603}.

\bibitem[{\citenamefont{Syrwid et~al.}(2017)\citenamefont{Syrwid, Zakrzewski,
  and Sacha}}]{Syrwid2017}
\bibinfo{author}{\bibfnamefont{A.}~\bibnamefont{Syrwid}},
  \bibinfo{author}{\bibfnamefont{J.}~\bibnamefont{Zakrzewski}},
  \bibnamefont{and} \bibinfo{author}{\bibfnamefont{K.}~\bibnamefont{Sacha}},
  \bibinfo{journal}{Phys. Rev. Lett.} \textbf{\bibinfo{volume}{119}},
  \bibinfo{pages}{250602} (\bibinfo{year}{2017}),
  \urlprefix\url{https://link.aps.org/doi/10.1103/PhysRevLett.119.250602}.

\bibitem[{\citenamefont{{Iemini} et~al.}(2017)\citenamefont{{Iemini},
  {Russomanno}, {Keeling}, {Schir{\`o}}, {Dalmonte}, and {Fazio}}}]{Iemini2017}
\bibinfo{author}{\bibfnamefont{F.}~\bibnamefont{{Iemini}}},
  \bibinfo{author}{\bibfnamefont{A.}~\bibnamefont{{Russomanno}}},
  \bibinfo{author}{\bibfnamefont{J.}~\bibnamefont{{Keeling}}},
  \bibinfo{author}{\bibfnamefont{M.}~\bibnamefont{{Schir{\`o}}}},
  \bibinfo{author}{\bibfnamefont{M.}~\bibnamefont{{Dalmonte}}},
  \bibnamefont{and} \bibinfo{author}{\bibfnamefont{R.}~\bibnamefont{{Fazio}}},
  \bibinfo{journal}{ArXiv e-prints}  (\bibinfo{year}{2017}),
  \eprint{1708.05014}.

\bibitem[{\citenamefont{Huang et~al.}(2018{\natexlab{a}})\citenamefont{Huang,
  Li, and Yin}}]{Huang2017a}
\bibinfo{author}{\bibfnamefont{Y.}~\bibnamefont{Huang}},
  \bibinfo{author}{\bibfnamefont{T.}~\bibnamefont{Li}}, \bibnamefont{and}
  \bibinfo{author}{\bibfnamefont{Z.-q.} \bibnamefont{Yin}},
  \bibinfo{journal}{Phys. Rev. A} \textbf{\bibinfo{volume}{97}},
  \bibinfo{pages}{012115} (\bibinfo{year}{2018}{\natexlab{a}}),
  \urlprefix\url{https://link.aps.org/doi/10.1103/PhysRevA.97.012115}.

\bibitem[{\citenamefont{{Prokof'ev} and {Svistunov}}(2017)}]{Prokofev2017}
\bibinfo{author}{\bibfnamefont{N.~V.} \bibnamefont{{Prokof'ev}}}
  \bibnamefont{and} \bibinfo{author}{\bibfnamefont{B.~V.}
  \bibnamefont{{Svistunov}}}, \bibinfo{journal}{ArXiv e-prints}
  (\bibinfo{year}{2017}), \eprint{1710.00721}.

\bibitem[{\citenamefont{Sacha}(2015{\natexlab{a}})}]{Sacha2015}
\bibinfo{author}{\bibfnamefont{K.}~\bibnamefont{Sacha}},
  \bibinfo{journal}{Phys. Rev. A} \textbf{\bibinfo{volume}{91}},
  \bibinfo{pages}{033617} (\bibinfo{year}{2015}{\natexlab{a}}),
  \urlprefix\url{http://link.aps.org/doi/10.1103/PhysRevA.91.033617}.

\bibitem[{\citenamefont{Khemani et~al.}(2016)\citenamefont{Khemani, Lazarides,
  Moessner, and Sondhi}}]{Khemani16}
\bibinfo{author}{\bibfnamefont{V.}~\bibnamefont{Khemani}},
  \bibinfo{author}{\bibfnamefont{A.}~\bibnamefont{Lazarides}},
  \bibinfo{author}{\bibfnamefont{R.}~\bibnamefont{Moessner}}, \bibnamefont{and}
  \bibinfo{author}{\bibfnamefont{S.~L.} \bibnamefont{Sondhi}},
  \bibinfo{journal}{Phys. Rev. Lett.} \textbf{\bibinfo{volume}{116}},
  \bibinfo{pages}{250401} (\bibinfo{year}{2016}),
  \urlprefix\url{http://link.aps.org/doi/10.1103/PhysRevLett.116.250401}.

\bibitem[{\citenamefont{Else et~al.}(2016)\citenamefont{Else, Bauer, and
  Nayak}}]{ElseFTC}
\bibinfo{author}{\bibfnamefont{D.~V.} \bibnamefont{Else}},
  \bibinfo{author}{\bibfnamefont{B.}~\bibnamefont{Bauer}}, \bibnamefont{and}
  \bibinfo{author}{\bibfnamefont{C.}~\bibnamefont{Nayak}},
  \bibinfo{journal}{Phys. Rev. Lett.} \textbf{\bibinfo{volume}{117}},
  \bibinfo{pages}{090402} (\bibinfo{year}{2016}),
  \urlprefix\url{http://link.aps.org/doi/10.1103/PhysRevLett.117.090402}.

\bibitem[{\citenamefont{Yao et~al.}(2017)\citenamefont{Yao, Potter, Potirniche,
  and Vishwanath}}]{Yao2017}
\bibinfo{author}{\bibfnamefont{N.~Y.} \bibnamefont{Yao}},
  \bibinfo{author}{\bibfnamefont{A.~C.} \bibnamefont{Potter}},
  \bibinfo{author}{\bibfnamefont{I.-D.} \bibnamefont{Potirniche}},
  \bibnamefont{and}
  \bibinfo{author}{\bibfnamefont{A.}~\bibnamefont{Vishwanath}},
  \bibinfo{journal}{Phys. Rev. Lett.} \textbf{\bibinfo{volume}{118}},
  \bibinfo{pages}{030401} (\bibinfo{year}{2017}),
  \urlprefix\url{http://link.aps.org/doi/10.1103/PhysRevLett.118.030401}.

\bibitem[{\citenamefont{Lazarides and Moessner}(2017)}]{Lazarides2017}
\bibinfo{author}{\bibfnamefont{A.}~\bibnamefont{Lazarides}} \bibnamefont{and}
  \bibinfo{author}{\bibfnamefont{R.}~\bibnamefont{Moessner}},
  \bibinfo{journal}{Phys. Rev. B} \textbf{\bibinfo{volume}{95}},
  \bibinfo{pages}{195135} (\bibinfo{year}{2017}),
  \urlprefix\url{https://link.aps.org/doi/10.1103/PhysRevB.95.195135}.

\bibitem[{\citenamefont{Russomanno et~al.}(2017)\citenamefont{Russomanno,
  Iemini, Dalmonte, and Fazio}}]{Russomanno2017}
\bibinfo{author}{\bibfnamefont{A.}~\bibnamefont{Russomanno}},
  \bibinfo{author}{\bibfnamefont{F.}~\bibnamefont{Iemini}},
  \bibinfo{author}{\bibfnamefont{M.}~\bibnamefont{Dalmonte}}, \bibnamefont{and}
  \bibinfo{author}{\bibfnamefont{R.}~\bibnamefont{Fazio}},
  \bibinfo{journal}{Phys. Rev. B} \textbf{\bibinfo{volume}{95}},
  \bibinfo{pages}{214307} (\bibinfo{year}{2017}),
  \urlprefix\url{https://link.aps.org/doi/10.1103/PhysRevB.95.214307}.

\bibitem[{\citenamefont{Zeng and Sheng}(2017)}]{Zeng2017}
\bibinfo{author}{\bibfnamefont{T.-S.} \bibnamefont{Zeng}} \bibnamefont{and}
  \bibinfo{author}{\bibfnamefont{D.~N.} \bibnamefont{Sheng}},
  \bibinfo{journal}{Phys. Rev. B} \textbf{\bibinfo{volume}{96}},
  \bibinfo{pages}{094202} (\bibinfo{year}{2017}),
  \urlprefix\url{https://link.aps.org/doi/10.1103/PhysRevB.96.094202}.

\bibitem[{\citenamefont{Nakatsugawa et~al.}(2017)\citenamefont{Nakatsugawa,
  Fujii, and Tanda}}]{Nakatsugawa2017}
\bibinfo{author}{\bibfnamefont{K.}~\bibnamefont{Nakatsugawa}},
  \bibinfo{author}{\bibfnamefont{T.}~\bibnamefont{Fujii}}, \bibnamefont{and}
  \bibinfo{author}{\bibfnamefont{S.}~\bibnamefont{Tanda}},
  \bibinfo{journal}{Phys. Rev. B} \textbf{\bibinfo{volume}{96}},
  \bibinfo{pages}{094308} (\bibinfo{year}{2017}),
  \urlprefix\url{https://link.aps.org/doi/10.1103/PhysRevB.96.094308}.

\bibitem[{\citenamefont{Ho et~al.}(2017)\citenamefont{Ho, Choi, Lukin, and
  Abanin}}]{Ho2017}
\bibinfo{author}{\bibfnamefont{W.~W.} \bibnamefont{Ho}},
  \bibinfo{author}{\bibfnamefont{S.}~\bibnamefont{Choi}},
  \bibinfo{author}{\bibfnamefont{M.~D.} \bibnamefont{Lukin}}, \bibnamefont{and}
  \bibinfo{author}{\bibfnamefont{D.~A.} \bibnamefont{Abanin}},
  \bibinfo{journal}{Phys. Rev. Lett.} \textbf{\bibinfo{volume}{119}},
  \bibinfo{pages}{010602} (\bibinfo{year}{2017}),
  \urlprefix\url{https://link.aps.org/doi/10.1103/PhysRevLett.119.010602}.

\bibitem[{\citenamefont{Huang et~al.}(2018{\natexlab{b}})\citenamefont{Huang,
  Wu, and Liu}}]{Huang2017}
\bibinfo{author}{\bibfnamefont{B.}~\bibnamefont{Huang}},
  \bibinfo{author}{\bibfnamefont{Y.-H.} \bibnamefont{Wu}}, \bibnamefont{and}
  \bibinfo{author}{\bibfnamefont{W.~V.} \bibnamefont{Liu}},
  \bibinfo{journal}{Phys. Rev. Lett.} \textbf{\bibinfo{volume}{120}},
  \bibinfo{pages}{110603} (\bibinfo{year}{2018}{\natexlab{b}}),
  \urlprefix\url{https://link.aps.org/doi/10.1103/PhysRevLett.120.110603}.

\bibitem[{\citenamefont{Gong et~al.}(2018)\citenamefont{Gong, Hamazaki, and
  Ueda}}]{Gong2017}
\bibinfo{author}{\bibfnamefont{Z.}~\bibnamefont{Gong}},
  \bibinfo{author}{\bibfnamefont{R.}~\bibnamefont{Hamazaki}}, \bibnamefont{and}
  \bibinfo{author}{\bibfnamefont{M.}~\bibnamefont{Ueda}},
  \bibinfo{journal}{Phys. Rev. Lett.} \textbf{\bibinfo{volume}{120}},
  \bibinfo{pages}{040404} (\bibinfo{year}{2018}),
  \urlprefix\url{https://link.aps.org/doi/10.1103/PhysRevLett.120.040404}.

\bibitem[{\citenamefont{Wang et~al.}(2018)\citenamefont{Wang, Xing, Carlo, and
  Poletti}}]{Wang2017}
\bibinfo{author}{\bibfnamefont{R.~R.~W.} \bibnamefont{Wang}},
  \bibinfo{author}{\bibfnamefont{B.}~\bibnamefont{Xing}},
  \bibinfo{author}{\bibfnamefont{G.~G.} \bibnamefont{Carlo}}, \bibnamefont{and}
  \bibinfo{author}{\bibfnamefont{D.}~\bibnamefont{Poletti}},
  \bibinfo{journal}{Phys. Rev. E} \textbf{\bibinfo{volume}{97}},
  \bibinfo{pages}{020202} (\bibinfo{year}{2018}),
  \urlprefix\url{https://link.aps.org/doi/10.1103/PhysRevE.97.020202}.

\bibitem[{\citenamefont{{Mizuta} et~al.}(2018)\citenamefont{{Mizuta},
  {Takasan}, {Nakagawa}, and {Kawakami}}}]{Mizuta2018}
\bibinfo{author}{\bibfnamefont{K.}~\bibnamefont{{Mizuta}}},
  \bibinfo{author}{\bibfnamefont{K.}~\bibnamefont{{Takasan}}},
  \bibinfo{author}{\bibfnamefont{M.}~\bibnamefont{{Nakagawa}}},
  \bibnamefont{and}
  \bibinfo{author}{\bibfnamefont{N.}~\bibnamefont{{Kawakami}}},
  \bibinfo{journal}{ArXiv e-prints}  (\bibinfo{year}{2018}),
  \eprint{1804.01291}.

\bibitem[{\citenamefont{Zhang et~al.}(2017)\citenamefont{Zhang, Hess,
  Kyprianidis, Becker, Lee, Smith, Pagano, Potirniche, Potter, Vishwanath
  et~al.}}]{Zhang2017}
\bibinfo{author}{\bibfnamefont{J.}~\bibnamefont{Zhang}},
  \bibinfo{author}{\bibfnamefont{P.~W.} \bibnamefont{Hess}},
  \bibinfo{author}{\bibfnamefont{A.}~\bibnamefont{Kyprianidis}},
  \bibinfo{author}{\bibfnamefont{P.}~\bibnamefont{Becker}},
  \bibinfo{author}{\bibfnamefont{A.}~\bibnamefont{Lee}},
  \bibinfo{author}{\bibfnamefont{J.}~\bibnamefont{Smith}},
  \bibinfo{author}{\bibfnamefont{G.}~\bibnamefont{Pagano}},
  \bibinfo{author}{\bibfnamefont{I.-D.} \bibnamefont{Potirniche}},
  \bibinfo{author}{\bibfnamefont{A.~C.} \bibnamefont{Potter}},
  \bibinfo{author}{\bibfnamefont{A.}~\bibnamefont{Vishwanath}},
  \bibnamefont{et~al.}, \bibinfo{journal}{Nature}
  \textbf{\bibinfo{volume}{543}}, \bibinfo{pages}{217} (\bibinfo{year}{2017}),
  ISSN \bibinfo{issn}{0028-0836}, \bibinfo{note}{letter},
  \urlprefix\url{http://dx.doi.org/10.1038/nature21413}.

\bibitem[{\citenamefont{Choi et~al.}(2017)\citenamefont{Choi, Choi, Landig,
  Kucsko, Zhou, Isoya, Jelezko, Onoda, Sumiya, Khemani et~al.}}]{Choi2017}
\bibinfo{author}{\bibfnamefont{S.}~\bibnamefont{Choi}},
  \bibinfo{author}{\bibfnamefont{J.}~\bibnamefont{Choi}},
  \bibinfo{author}{\bibfnamefont{R.}~\bibnamefont{Landig}},
  \bibinfo{author}{\bibfnamefont{G.}~\bibnamefont{Kucsko}},
  \bibinfo{author}{\bibfnamefont{H.}~\bibnamefont{Zhou}},
  \bibinfo{author}{\bibfnamefont{J.}~\bibnamefont{Isoya}},
  \bibinfo{author}{\bibfnamefont{F.}~\bibnamefont{Jelezko}},
  \bibinfo{author}{\bibfnamefont{S.}~\bibnamefont{Onoda}},
  \bibinfo{author}{\bibfnamefont{H.}~\bibnamefont{Sumiya}},
  \bibinfo{author}{\bibfnamefont{V.}~\bibnamefont{Khemani}},
  \bibnamefont{et~al.}, \bibinfo{journal}{Nature}
  \textbf{\bibinfo{volume}{543}}, \bibinfo{pages}{221} (\bibinfo{year}{2017}),
  ISSN \bibinfo{issn}{0028-0836}, \bibinfo{note}{letter},
  \urlprefix\url{http://dx.doi.org/10.1038/nature21426}.

\bibitem[{\citenamefont{Nayak}(2017)}]{Nayak2017}
\bibinfo{author}{\bibfnamefont{C.}~\bibnamefont{Nayak}},
  \bibinfo{journal}{Nature} \textbf{\bibinfo{volume}{543}},
  \bibinfo{pages}{185} (\bibinfo{year}{2017}), ISSN \bibinfo{issn}{0028-0836},
  \bibinfo{note}{news \& Views},
  \urlprefix\url{http://dx.doi.org/10.1038/543185a}.

\bibitem[{\citenamefont{Pal et~al.}(2018)\citenamefont{Pal, Nishad, Mahesh, and
  Sreejith}}]{Pal2018}
\bibinfo{author}{\bibfnamefont{S.}~\bibnamefont{Pal}},
  \bibinfo{author}{\bibfnamefont{N.}~\bibnamefont{Nishad}},
  \bibinfo{author}{\bibfnamefont{T.~S.} \bibnamefont{Mahesh}},
  \bibnamefont{and} \bibinfo{author}{\bibfnamefont{G.~J.}
  \bibnamefont{Sreejith}}, \bibinfo{journal}{Phys. Rev. Lett.}
  \textbf{\bibinfo{volume}{120}}, \bibinfo{pages}{180602}
  (\bibinfo{year}{2018}),
  \urlprefix\url{https://link.aps.org/doi/10.1103/PhysRevLett.120.180602}.

\bibitem[{\citenamefont{Rovny et~al.}(2018{\natexlab{a}})\citenamefont{Rovny,
  Blum, and Barrett}}]{Rovny2018}
\bibinfo{author}{\bibfnamefont{J.}~\bibnamefont{Rovny}},
  \bibinfo{author}{\bibfnamefont{R.~L.} \bibnamefont{Blum}}, \bibnamefont{and}
  \bibinfo{author}{\bibfnamefont{S.~E.} \bibnamefont{Barrett}},
  \bibinfo{journal}{Phys. Rev. Lett.} \textbf{\bibinfo{volume}{120}},
  \bibinfo{pages}{180603} (\bibinfo{year}{2018}{\natexlab{a}}),
  \urlprefix\url{https://link.aps.org/doi/10.1103/PhysRevLett.120.180603}.

\bibitem[{\citenamefont{Rovny et~al.}(2018{\natexlab{b}})\citenamefont{Rovny,
  Blum, and Barrett}}]{Rovny2018a}
\bibinfo{author}{\bibfnamefont{J.}~\bibnamefont{Rovny}},
  \bibinfo{author}{\bibfnamefont{R.~L.} \bibnamefont{Blum}}, \bibnamefont{and}
  \bibinfo{author}{\bibfnamefont{S.~E.} \bibnamefont{Barrett}},
  \bibinfo{journal}{Phys. Rev. B} \textbf{\bibinfo{volume}{97}},
  \bibinfo{pages}{184301} (\bibinfo{year}{2018}{\natexlab{b}}),
  \urlprefix\url{https://link.aps.org/doi/10.1103/PhysRevB.97.184301}.

\bibitem[{\citenamefont{Autti et~al.}(2018)\citenamefont{Autti, Eltsov, and
  Volovik}}]{Autti2018}
\bibinfo{author}{\bibfnamefont{S.}~\bibnamefont{Autti}},
  \bibinfo{author}{\bibfnamefont{V.~B.} \bibnamefont{Eltsov}},
  \bibnamefont{and} \bibinfo{author}{\bibfnamefont{G.~E.}
  \bibnamefont{Volovik}}, \bibinfo{journal}{Phys. Rev. Lett.}
  \textbf{\bibinfo{volume}{120}}, \bibinfo{pages}{215301}
  (\bibinfo{year}{2018}),
  \urlprefix\url{https://link.aps.org/doi/10.1103/PhysRevLett.120.215301}.

\bibitem[{\citenamefont{Kim et~al.}(2006)\citenamefont{Kim, Heo, Lee, Jang,
  Noh, Kim, and Jhe}}]{Kim2006}
\bibinfo{author}{\bibfnamefont{K.}~\bibnamefont{Kim}},
  \bibinfo{author}{\bibfnamefont{M.-S.} \bibnamefont{Heo}},
  \bibinfo{author}{\bibfnamefont{K.-H.} \bibnamefont{Lee}},
  \bibinfo{author}{\bibfnamefont{K.}~\bibnamefont{Jang}},
  \bibinfo{author}{\bibfnamefont{H.-R.} \bibnamefont{Noh}},
  \bibinfo{author}{\bibfnamefont{D.}~\bibnamefont{Kim}}, \bibnamefont{and}
  \bibinfo{author}{\bibfnamefont{W.}~\bibnamefont{Jhe}},
  \bibinfo{journal}{Phys. Rev. Lett.} \textbf{\bibinfo{volume}{96}},
  \bibinfo{pages}{150601} (\bibinfo{year}{2006}),
  \urlprefix\url{https://link.aps.org/doi/10.1103/PhysRevLett.96.150601}.

\bibitem[{\citenamefont{Heo et~al.}(2010)\citenamefont{Heo, Kim, Kim, Moon,
  Lee, Noh, Dykman, and Jhe}}]{Heo2010}
\bibinfo{author}{\bibfnamefont{M.-S.} \bibnamefont{Heo}},
  \bibinfo{author}{\bibfnamefont{Y.}~\bibnamefont{Kim}},
  \bibinfo{author}{\bibfnamefont{K.}~\bibnamefont{Kim}},
  \bibinfo{author}{\bibfnamefont{G.}~\bibnamefont{Moon}},
  \bibinfo{author}{\bibfnamefont{J.}~\bibnamefont{Lee}},
  \bibinfo{author}{\bibfnamefont{H.-R.} \bibnamefont{Noh}},
  \bibinfo{author}{\bibfnamefont{M.~I.} \bibnamefont{Dykman}},
  \bibnamefont{and} \bibinfo{author}{\bibfnamefont{W.}~\bibnamefont{Jhe}},
  \bibinfo{journal}{Phys. Rev. E} \textbf{\bibinfo{volume}{82}},
  \bibinfo{pages}{031134} (\bibinfo{year}{2010}),
  \urlprefix\url{https://link.aps.org/doi/10.1103/PhysRevE.82.031134}.

\bibitem[{\citenamefont{Zi\'n et~al.}(2008)\citenamefont{Zi\'n, Chwede\'nczuk,
  Ole\'s, Sacha, and Trippenbach}}]{Zin2008}
\bibinfo{author}{\bibfnamefont{P.}~\bibnamefont{Zi\'n}},
  \bibinfo{author}{\bibfnamefont{J.}~\bibnamefont{Chwede\'nczuk}},
  \bibinfo{author}{\bibfnamefont{B.}~\bibnamefont{Ole\'s}},
  \bibinfo{author}{\bibfnamefont{K.}~\bibnamefont{Sacha}}, \bibnamefont{and}
  \bibinfo{author}{\bibfnamefont{M.}~\bibnamefont{Trippenbach}},
  \bibinfo{journal}{EPL (Europhysics Letters)} \textbf{\bibinfo{volume}{83}},
  \bibinfo{pages}{64007} (\bibinfo{year}{2008}),
  \urlprefix\url{http://stacks.iop.org/0295-5075/83/i=6/a=64007}.

\bibitem[{\citenamefont{Kosior and Sacha}(2018)}]{Kosior2017}
\bibinfo{author}{\bibfnamefont{A.}~\bibnamefont{Kosior}} \bibnamefont{and}
  \bibinfo{author}{\bibfnamefont{K.}~\bibnamefont{Sacha}},
  \bibinfo{journal}{Phys. Rev. A} \textbf{\bibinfo{volume}{97}},
  \bibinfo{pages}{053621} (\bibinfo{year}{2018}),
  \urlprefix\url{https://link.aps.org/doi/10.1103/PhysRevA.97.053621}.

\bibitem[{\citenamefont{Heyl et~al.}(2013)\citenamefont{Heyl, Polkovnikov, and
  Kehrein}}]{Heyl2013}
\bibinfo{author}{\bibfnamefont{M.}~\bibnamefont{Heyl}},
  \bibinfo{author}{\bibfnamefont{A.}~\bibnamefont{Polkovnikov}},
  \bibnamefont{and} \bibinfo{author}{\bibfnamefont{S.}~\bibnamefont{Kehrein}},
  \bibinfo{journal}{Phys. Rev. Lett.} \textbf{\bibinfo{volume}{110}},
  \bibinfo{pages}{135704} (\bibinfo{year}{2013}),
  \urlprefix\url{https://link.aps.org/doi/10.1103/PhysRevLett.110.135704}.

\bibitem[{\citenamefont{Jurcevic et~al.}(2017)\citenamefont{Jurcevic, Shen,
  Hauke, Maier, Brydges, Hempel, Lanyon, Heyl, Blatt, and Roos}}]{Jurcevic2017}
\bibinfo{author}{\bibfnamefont{P.}~\bibnamefont{Jurcevic}},
  \bibinfo{author}{\bibfnamefont{H.}~\bibnamefont{Shen}},
  \bibinfo{author}{\bibfnamefont{P.}~\bibnamefont{Hauke}},
  \bibinfo{author}{\bibfnamefont{C.}~\bibnamefont{Maier}},
  \bibinfo{author}{\bibfnamefont{T.}~\bibnamefont{Brydges}},
  \bibinfo{author}{\bibfnamefont{C.}~\bibnamefont{Hempel}},
  \bibinfo{author}{\bibfnamefont{B.~P.} \bibnamefont{Lanyon}},
  \bibinfo{author}{\bibfnamefont{M.}~\bibnamefont{Heyl}},
  \bibinfo{author}{\bibfnamefont{R.}~\bibnamefont{Blatt}}, \bibnamefont{and}
  \bibinfo{author}{\bibfnamefont{C.~F.} \bibnamefont{Roos}},
  \bibinfo{journal}{Phys. Rev. Lett.} \textbf{\bibinfo{volume}{119}},
  \bibinfo{pages}{080501} (\bibinfo{year}{2017}),
  \urlprefix\url{https://link.aps.org/doi/10.1103/PhysRevLett.119.080501}.

\bibitem[{\citenamefont{Fl\"aschner et~al.}(2018)\citenamefont{Fl\"aschner,
  Vogel, Tarnowski, Rem, L\"uhmann, Heyl, Budich, Mathey, Sengstock, and
  Weitenberg}}]{Flaschner2018}
\bibinfo{author}{\bibfnamefont{N.}~\bibnamefont{Fl\"aschner}},
  \bibinfo{author}{\bibfnamefont{D.}~\bibnamefont{Vogel}},
  \bibinfo{author}{\bibfnamefont{M.}~\bibnamefont{Tarnowski}},
  \bibinfo{author}{\bibfnamefont{B.~S.} \bibnamefont{Rem}},
  \bibinfo{author}{\bibfnamefont{D.-S.} \bibnamefont{L\"uhmann}},
  \bibinfo{author}{\bibfnamefont{M.}~\bibnamefont{Heyl}},
  \bibinfo{author}{\bibfnamefont{J.~C.} \bibnamefont{Budich}},
  \bibinfo{author}{\bibfnamefont{L.}~\bibnamefont{Mathey}},
  \bibinfo{author}{\bibfnamefont{K.}~\bibnamefont{Sengstock}},
  \bibnamefont{and}
  \bibinfo{author}{\bibfnamefont{C.}~\bibnamefont{Weitenberg}},
  \bibinfo{journal}{Nature Physics} \textbf{\bibinfo{volume}{14}},
  \bibinfo{pages}{265} (\bibinfo{year}{2018}).

\bibitem[{\citenamefont{{Heyl}}(2018)}]{Heyl2018rev}
\bibinfo{author}{\bibfnamefont{M.}~\bibnamefont{{Heyl}}},
  \bibinfo{journal}{Rep. Prog. Phys.} \textbf{\bibinfo{volume}{81}},
  \bibinfo{pages}{054001} (\bibinfo{year}{2018}).

\bibitem[{\citenamefont{Steane et~al.}(1995)\citenamefont{Steane, Szriftgiser,
  Desbiolles, and Dalibard}}]{Steane95}
\bibinfo{author}{\bibfnamefont{A.}~\bibnamefont{Steane}},
  \bibinfo{author}{\bibfnamefont{P.}~\bibnamefont{Szriftgiser}},
  \bibinfo{author}{\bibfnamefont{P.}~\bibnamefont{Desbiolles}},
  \bibnamefont{and} \bibinfo{author}{\bibfnamefont{J.}~\bibnamefont{Dalibard}},
  \bibinfo{journal}{Phys. Rev. Lett.} \textbf{\bibinfo{volume}{74}},
  \bibinfo{pages}{4972} (\bibinfo{year}{1995}),
  \urlprefix\url{http://link.aps.org/doi/10.1103/PhysRevLett.74.4972}.

\bibitem[{\citenamefont{Roach et~al.}(1995)\citenamefont{Roach, Abele, Boshier,
  Grossman, Zetie, and Hinds}}]{Roach1995}
\bibinfo{author}{\bibfnamefont{T.~M.} \bibnamefont{Roach}},
  \bibinfo{author}{\bibfnamefont{H.}~\bibnamefont{Abele}},
  \bibinfo{author}{\bibfnamefont{M.~G.} \bibnamefont{Boshier}},
  \bibinfo{author}{\bibfnamefont{H.~L.} \bibnamefont{Grossman}},
  \bibinfo{author}{\bibfnamefont{K.~P.} \bibnamefont{Zetie}}, \bibnamefont{and}
  \bibinfo{author}{\bibfnamefont{E.~A.} \bibnamefont{Hinds}},
  \bibinfo{journal}{Phys. Rev. Lett.} \textbf{\bibinfo{volume}{75}},
  \bibinfo{pages}{629} (\bibinfo{year}{1995}),
  \urlprefix\url{https://link.aps.org/doi/10.1103/PhysRevLett.75.629}.

\bibitem[{\citenamefont{Sidorov et~al.}(1996)\citenamefont{Sidorov, McLean,
  Rowlands, Lau, Murphy, Walkiewicz, Opat, and Hannaford}}]{Sidorov1996}
\bibinfo{author}{\bibfnamefont{A.~I.} \bibnamefont{Sidorov}},
  \bibinfo{author}{\bibfnamefont{R.~J.} \bibnamefont{McLean}},
  \bibinfo{author}{\bibfnamefont{W.~J.} \bibnamefont{Rowlands}},
  \bibinfo{author}{\bibfnamefont{D.~C.} \bibnamefont{Lau}},
  \bibinfo{author}{\bibfnamefont{J.~E.} \bibnamefont{Murphy}},
  \bibinfo{author}{\bibfnamefont{M.}~\bibnamefont{Walkiewicz}},
  \bibinfo{author}{\bibfnamefont{G.~I.} \bibnamefont{Opat}}, \bibnamefont{and}
  \bibinfo{author}{\bibfnamefont{P.}~\bibnamefont{Hannaford}},
  \bibinfo{journal}{Quantum and Semiclassical Optics: Journal of the European
  Optical Society Part B} \textbf{\bibinfo{volume}{8}}, \bibinfo{pages}{713}
  (\bibinfo{year}{1996}),
  \urlprefix\url{http://stacks.iop.org/1355-5111/8/i=3/a=030}.

\bibitem[{\citenamefont{Westbrook et~al.}(1998)\citenamefont{Westbrook,
  Westbrook, Landragin, Labeyrie, Cognet, Savalli, Horvath, Aspect, Hendel,
  Moelmer et~al.}}]{Westbrook1998}
\bibinfo{author}{\bibfnamefont{N.}~\bibnamefont{Westbrook}},
  \bibinfo{author}{\bibfnamefont{C.~I.} \bibnamefont{Westbrook}},
  \bibinfo{author}{\bibfnamefont{A.}~\bibnamefont{Landragin}},
  \bibinfo{author}{\bibfnamefont{G.}~\bibnamefont{Labeyrie}},
  \bibinfo{author}{\bibfnamefont{L.}~\bibnamefont{Cognet}},
  \bibinfo{author}{\bibfnamefont{V.}~\bibnamefont{Savalli}},
  \bibinfo{author}{\bibfnamefont{G.}~\bibnamefont{Horvath}},
  \bibinfo{author}{\bibfnamefont{A.}~\bibnamefont{Aspect}},
  \bibinfo{author}{\bibfnamefont{C.}~\bibnamefont{Hendel}},
  \bibinfo{author}{\bibfnamefont{K.}~\bibnamefont{Moelmer}},
  \bibnamefont{et~al.}, \bibinfo{journal}{Phys. Scr. T}
  \textbf{\bibinfo{volume}{78}}, \bibinfo{pages}{7} (\bibinfo{year}{1998}).

\bibitem[{\citenamefont{Lau et~al.}(1999)\citenamefont{Lau, Sidorov, Opat,
  McLean, Rowlands, and Hannaford}}]{Lau1999}
\bibinfo{author}{\bibfnamefont{D.~C.} \bibnamefont{Lau}},
  \bibinfo{author}{\bibfnamefont{A.~I.} \bibnamefont{Sidorov}},
  \bibinfo{author}{\bibfnamefont{G.~I.} \bibnamefont{Opat}},
  \bibinfo{author}{\bibfnamefont{R.~J.} \bibnamefont{McLean}},
  \bibinfo{author}{\bibfnamefont{W.~J.} \bibnamefont{Rowlands}},
  \bibnamefont{and}
  \bibinfo{author}{\bibfnamefont{P.}~\bibnamefont{Hannaford}},
  \bibinfo{journal}{Eur. Phys. J. D} \textbf{\bibinfo{volume}{5}},
  \bibinfo{pages}{193} (\bibinfo{year}{1999}),
  \urlprefix\url{https://doi.org/10.1007/s100530050244}.

\bibitem[{\citenamefont{Bongs et~al.}(1999)\citenamefont{Bongs, Burger, Birkl,
  Sengstock, Ertmer, Rz\c{a}\.zewski, Sanpera, and Lewenstein}}]{Bongs1999}
\bibinfo{author}{\bibfnamefont{K.}~\bibnamefont{Bongs}},
  \bibinfo{author}{\bibfnamefont{S.}~\bibnamefont{Burger}},
  \bibinfo{author}{\bibfnamefont{G.}~\bibnamefont{Birkl}},
  \bibinfo{author}{\bibfnamefont{K.}~\bibnamefont{Sengstock}},
  \bibinfo{author}{\bibfnamefont{W.}~\bibnamefont{Ertmer}},
  \bibinfo{author}{\bibfnamefont{K.}~\bibnamefont{Rz\c{a}\.zewski}},
  \bibinfo{author}{\bibfnamefont{A.}~\bibnamefont{Sanpera}}, \bibnamefont{and}
  \bibinfo{author}{\bibfnamefont{M.}~\bibnamefont{Lewenstein}},
  \bibinfo{journal}{Phys. Rev. Lett.} \textbf{\bibinfo{volume}{83}},
  \bibinfo{pages}{3577} (\bibinfo{year}{1999}),
  \urlprefix\url{https://link.aps.org/doi/10.1103/PhysRevLett.83.3577}.

\bibitem[{\citenamefont{Sidorov et~al.}(2002)\citenamefont{Sidorov, McLean,
  Scharnberg, Gough, Davis, Sexton, Opat, and Hannaford}}]{Sidorov2002}
\bibinfo{author}{\bibfnamefont{A.}~\bibnamefont{Sidorov}},
  \bibinfo{author}{\bibfnamefont{R.}~\bibnamefont{McLean}},
  \bibinfo{author}{\bibfnamefont{F.}~\bibnamefont{Scharnberg}},
  \bibinfo{author}{\bibfnamefont{D.}~\bibnamefont{Gough}},
  \bibinfo{author}{\bibfnamefont{T.}~\bibnamefont{Davis}},
  \bibinfo{author}{\bibfnamefont{B.}~\bibnamefont{Sexton}},
  \bibinfo{author}{\bibfnamefont{G.}~\bibnamefont{Opat}}, \bibnamefont{and}
  \bibinfo{author}{\bibfnamefont{P.}~\bibnamefont{Hannaford}},
  \bibinfo{journal}{Acta Phys. Pol. B} \textbf{\bibinfo{volume}{33}},
  \bibinfo{pages}{2137} (\bibinfo{year}{2002}).

\bibitem[{\citenamefont{Fiutowski et~al.}(2013)\citenamefont{Fiutowski,
  Bartoszek-Bober, Dohnalik, and Kawalec}}]{Fiutowski2013}
\bibinfo{author}{\bibfnamefont{J.}~\bibnamefont{Fiutowski}},
  \bibinfo{author}{\bibfnamefont{D.}~\bibnamefont{Bartoszek-Bober}},
  \bibinfo{author}{\bibfnamefont{T.}~\bibnamefont{Dohnalik}}, \bibnamefont{and}
  \bibinfo{author}{\bibfnamefont{T.}~\bibnamefont{Kawalec}},
  \bibinfo{journal}{Opt. Commun.} \textbf{\bibinfo{volume}{297}},
  \bibinfo{pages}{59} (\bibinfo{year}{2013}).

\bibitem[{\citenamefont{Kawalec et~al.}(2014)\citenamefont{Kawalec,
  Bartoszek-Bober, Pana\'{s}, Fiutowski, P{\l}awecka, and
  Rubahn}}]{Kawalec2014}
\bibinfo{author}{\bibfnamefont{T.}~\bibnamefont{Kawalec}},
  \bibinfo{author}{\bibfnamefont{D.}~\bibnamefont{Bartoszek-Bober}},
  \bibinfo{author}{\bibfnamefont{R.}~\bibnamefont{Pana\'{s}}},
  \bibinfo{author}{\bibfnamefont{J.}~\bibnamefont{Fiutowski}},
  \bibinfo{author}{\bibfnamefont{A.}~\bibnamefont{P{\l}awecka}},
  \bibnamefont{and} \bibinfo{author}{\bibfnamefont{H.-G.}
  \bibnamefont{Rubahn}}, \bibinfo{journal}{Opt. Lett.}
  \textbf{\bibinfo{volume}{39}}, \bibinfo{pages}{2932} (\bibinfo{year}{2014}),
  \urlprefix\url{http://ol.osa.org/abstract.cfm?URI=ol-39-10-2932}.

\bibitem[{\citenamefont{Guo et~al.}(2013)\citenamefont{Guo, Marthaler, and
  Sch\"on}}]{Guo2013}
\bibinfo{author}{\bibfnamefont{L.}~\bibnamefont{Guo}},
  \bibinfo{author}{\bibfnamefont{M.}~\bibnamefont{Marthaler}},
  \bibnamefont{and} \bibinfo{author}{\bibfnamefont{G.}~\bibnamefont{Sch\"on}},
  \bibinfo{journal}{Phys. Rev. Lett.} \textbf{\bibinfo{volume}{111}},
  \bibinfo{pages}{205303} (\bibinfo{year}{2013}),
  \urlprefix\url{https://link.aps.org/doi/10.1103/PhysRevLett.111.205303}.

\bibitem[{\citenamefont{Sacha}(2015{\natexlab{b}})}]{Sacha15a}
\bibinfo{author}{\bibfnamefont{K.}~\bibnamefont{Sacha}}, \bibinfo{journal}{Sci.
  Rep.} \textbf{\bibinfo{volume}{5}}, \bibinfo{pages}{10787}
  (\bibinfo{year}{2015}{\natexlab{b}}),
  \urlprefix\url{https://www.nature.com/articles/srep10787}.

\bibitem[{\citenamefont{Giergiel et~al.}(2018)\citenamefont{Giergiel,
  Miroszewski, and Sacha}}]{Giergiel2018}
\bibinfo{author}{\bibfnamefont{K.}~\bibnamefont{Giergiel}},
  \bibinfo{author}{\bibfnamefont{A.}~\bibnamefont{Miroszewski}},
  \bibnamefont{and} \bibinfo{author}{\bibfnamefont{K.}~\bibnamefont{Sacha}},
  \bibinfo{journal}{Phys. Rev. Lett.} \textbf{\bibinfo{volume}{120}},
  \bibinfo{pages}{140401} (\bibinfo{year}{2018}),
  \urlprefix\url{https://link.aps.org/doi/10.1103/PhysRevLett.120.140401}.

\bibitem[{\citenamefont{Sacha and Delande}(2016)}]{sacha16}
\bibinfo{author}{\bibfnamefont{K.}~\bibnamefont{Sacha}} \bibnamefont{and}
  \bibinfo{author}{\bibfnamefont{D.}~\bibnamefont{Delande}},
  \bibinfo{journal}{Phys. Rev. A} \textbf{\bibinfo{volume}{94}},
  \bibinfo{pages}{023633} (\bibinfo{year}{2016}),
  \urlprefix\url{http://link.aps.org/doi/10.1103/PhysRevA.94.023633}.

\bibitem[{\citenamefont{Giergiel and Sacha}(2017)}]{Giergiel2017}
\bibinfo{author}{\bibfnamefont{K.}~\bibnamefont{Giergiel}} \bibnamefont{and}
  \bibinfo{author}{\bibfnamefont{K.}~\bibnamefont{Sacha}},
  \bibinfo{journal}{Phys. Rev. A} \textbf{\bibinfo{volume}{95}},
  \bibinfo{pages}{063402} (\bibinfo{year}{2017}),
  \urlprefix\url{https://link.aps.org/doi/10.1103/PhysRevA.95.063402}.

\bibitem[{\citenamefont{Delande et~al.}(2017)\citenamefont{Delande,
  Morales-Molina, and Sacha}}]{delande17}
\bibinfo{author}{\bibfnamefont{D.}~\bibnamefont{Delande}},
  \bibinfo{author}{\bibfnamefont{L.}~\bibnamefont{Morales-Molina}},
  \bibnamefont{and} \bibinfo{author}{\bibfnamefont{K.}~\bibnamefont{Sacha}},
  \bibinfo{journal}{Phys. Rev. Lett.} \textbf{\bibinfo{volume}{119}},
  \bibinfo{pages}{230404} (\bibinfo{year}{2017}),
  \urlprefix\url{https://link.aps.org/doi/10.1103/PhysRevLett.119.230404}.

\bibitem[{\citenamefont{Buchleitner et~al.}(2002)\citenamefont{Buchleitner,
  Delande, and Zakrzewski}}]{Buchleitner2002}
\bibinfo{author}{\bibfnamefont{A.}~\bibnamefont{Buchleitner}},
  \bibinfo{author}{\bibfnamefont{D.}~\bibnamefont{Delande}}, \bibnamefont{and}
  \bibinfo{author}{\bibfnamefont{J.}~\bibnamefont{Zakrzewski}},
  \bibinfo{journal}{Physics reports} \textbf{\bibinfo{volume}{368}},
  \bibinfo{pages}{409} (\bibinfo{year}{2002}),
  \urlprefix\url{http://www.sciencedirect.com/science/article/pii/S0370157302002703}.

\bibitem[{\citenamefont{Shirley}(1965)}]{Shirley1965}
\bibinfo{author}{\bibfnamefont{J.~H.} \bibnamefont{Shirley}},
  \bibinfo{journal}{Phys. Rev.} \textbf{\bibinfo{volume}{138}},
  \bibinfo{pages}{B979} (\bibinfo{year}{1965}),
  \urlprefix\url{https://link.aps.org/doi/10.1103/PhysRev.138.B979}.

\bibitem[{\citenamefont{Berman and Zaslavsky}(1977)}]{Berman1977}
\bibinfo{author}{\bibfnamefont{G.}~\bibnamefont{Berman}} \bibnamefont{and}
  \bibinfo{author}{\bibfnamefont{G.}~\bibnamefont{Zaslavsky}},
  \bibinfo{journal}{Physics Letters A} \textbf{\bibinfo{volume}{61}},
  \bibinfo{pages}{295} (\bibinfo{year}{1977}), ISSN \bibinfo{issn}{0375-9601},
  \urlprefix\url{http://www.sciencedirect.com/science/article/pii/0375960177906181}.

\bibitem[{\citenamefont{Lichtenberg and Lieberman}(1992)}]{Lichtenberg1992}
\bibinfo{author}{\bibfnamefont{A.}~\bibnamefont{Lichtenberg}} \bibnamefont{and}
  \bibinfo{author}{\bibfnamefont{M.}~\bibnamefont{Lieberman}},
  \emph{\bibinfo{title}{Regular and chaotic dynamics}}, Applied mathematical
  sciences (\bibinfo{publisher}{Springer-Verlag}, \bibinfo{year}{1992}), ISBN
  \bibinfo{isbn}{9783540977452},
  \urlprefix\url{https://books.google.pl/books?id=2ssPAQAAMAAJ}.

\bibitem[{\citenamefont{Guo and Marthaler}(2016)}]{Guo2016}
\bibinfo{author}{\bibfnamefont{L.}~\bibnamefont{Guo}} \bibnamefont{and}
  \bibinfo{author}{\bibfnamefont{M.}~\bibnamefont{Marthaler}},
  \bibinfo{journal}{New Journal of Physics} \textbf{\bibinfo{volume}{18}},
  \bibinfo{pages}{023006} (\bibinfo{year}{2016}),
  \urlprefix\url{http://stacks.iop.org/1367-2630/18/i=2/a=023006}.

\bibitem[{\citenamefont{Guo et~al.}(2016)\citenamefont{Guo, Liu, and
  Marthaler}}]{Guo2016a}
\bibinfo{author}{\bibfnamefont{L.}~\bibnamefont{Guo}},
  \bibinfo{author}{\bibfnamefont{M.}~\bibnamefont{Liu}}, \bibnamefont{and}
  \bibinfo{author}{\bibfnamefont{M.}~\bibnamefont{Marthaler}},
  \bibinfo{journal}{Phys. Rev. A} \textbf{\bibinfo{volume}{93}},
  \bibinfo{pages}{053616} (\bibinfo{year}{2016}),
  \urlprefix\url{https://link.aps.org/doi/10.1103/PhysRevA.93.053616}.

\bibitem[{\citenamefont{Liang et~al.}(2018)\citenamefont{Liang, Marthaler, and
  Lingzhen}}]{Liang2017}
\bibinfo{author}{\bibfnamefont{P.}~\bibnamefont{Liang}},
  \bibinfo{author}{\bibfnamefont{M.}~\bibnamefont{Marthaler}},
  \bibnamefont{and} \bibinfo{author}{\bibfnamefont{G.}~\bibnamefont{Lingzhen}},
  \bibinfo{journal}{New Journal of Physics} \textbf{\bibinfo{volume}{20}},
  \bibinfo{pages}{023043} (\bibinfo{year}{2018}), ISSN
  \bibinfo{issn}{1367-2630},
  \urlprefix\url{http://stacks.iop.org/1367-2630/20/i=2/a=023043}.

\bibitem[{\citenamefont{Mierzejewski et~al.}(2017)\citenamefont{Mierzejewski,
  Giergiel, and Sacha}}]{Mierzejewski2017}
\bibinfo{author}{\bibfnamefont{M.}~\bibnamefont{Mierzejewski}},
  \bibinfo{author}{\bibfnamefont{K.}~\bibnamefont{Giergiel}}, \bibnamefont{and}
  \bibinfo{author}{\bibfnamefont{K.}~\bibnamefont{Sacha}},
  \bibinfo{journal}{Phys. Rev. B} \textbf{\bibinfo{volume}{96}},
  \bibinfo{pages}{140201} (\bibinfo{year}{2017}),
  \urlprefix\url{https://link.aps.org/doi/10.1103/PhysRevB.96.140201}.

\bibitem[{\citenamefont{Dutta et~al.}(2015)\citenamefont{Dutta, Gajda, Hauke,
  Lewenstein, L\"uhmann, Malomed, Sowi\'nski, and Zakrzewski}}]{Dutta2015}
\bibinfo{author}{\bibfnamefont{O.}~\bibnamefont{Dutta}},
  \bibinfo{author}{\bibfnamefont{M.}~\bibnamefont{Gajda}},
  \bibinfo{author}{\bibfnamefont{P.}~\bibnamefont{Hauke}},
  \bibinfo{author}{\bibfnamefont{M.}~\bibnamefont{Lewenstein}},
  \bibinfo{author}{\bibfnamefont{D.-S.} \bibnamefont{L\"uhmann}},
  \bibinfo{author}{\bibfnamefont{B.~A.} \bibnamefont{Malomed}},
  \bibinfo{author}{\bibfnamefont{T.}~\bibnamefont{Sowi\'nski}},
  \bibnamefont{and}
  \bibinfo{author}{\bibfnamefont{J.}~\bibnamefont{Zakrzewski}},
  \bibinfo{journal}{Reports on Progress in Physics}
  \textbf{\bibinfo{volume}{78}}, \bibinfo{pages}{066001}
  (\bibinfo{year}{2015}), ISSN \bibinfo{issn}{0034-4885},
  \urlprefix\url{http://stacks.iop.org/0034-4885/78/i=6/a=066001}.

\bibitem[{\citenamefont{Pethick and Smith}(2002)}]{Pethick2002}
\bibinfo{author}{\bibfnamefont{C.}~\bibnamefont{Pethick}} \bibnamefont{and}
  \bibinfo{author}{\bibfnamefont{H.}~\bibnamefont{Smith}},
  \emph{\bibinfo{title}{{Bose-Eistein condensation in dilute gases}}}
  (\bibinfo{publisher}{{Cambridge University Press}},
  \bibinfo{address}{{Cambridge, England}}, \bibinfo{year}{2002}).

\bibitem[{\citenamefont{Castin}(2001)}]{Castin_LesHouches}
\bibinfo{author}{\bibfnamefont{Y.}~\bibnamefont{Castin}}, in
  \emph{\bibinfo{booktitle}{Coherent atomic matter waves}}, edited by
  \bibinfo{editor}{\bibfnamefont{R.}~\bibnamefont{Kaiser}},
  \bibinfo{editor}{\bibfnamefont{C.}~\bibnamefont{Westbrook}},
  \bibnamefont{and} \bibinfo{editor}{\bibfnamefont{F.}~\bibnamefont{David}}
  (\bibinfo{publisher}{Springer Berlin Heidelberg}, \bibinfo{address}{Berlin,
  Heidelberg}, \bibinfo{year}{2001}), pp. \bibinfo{pages}{1--136}, ISBN
  \bibinfo{isbn}{978-3-540-45338-3}.

\bibitem[{\citenamefont{Saito and Ueda}(2001{\natexlab{a}})}]{Saito2001}
\bibinfo{author}{\bibfnamefont{H.}~\bibnamefont{Saito}} \bibnamefont{and}
  \bibinfo{author}{\bibfnamefont{M.}~\bibnamefont{Ueda}},
  \bibinfo{journal}{Phys. Rev. Lett.} \textbf{\bibinfo{volume}{86}},
  \bibinfo{pages}{1406} (\bibinfo{year}{2001}{\natexlab{a}}),
  \urlprefix\url{https://link.aps.org/doi/10.1103/PhysRevLett.86.1406}.

\bibitem[{\citenamefont{Saito and Ueda}(2001{\natexlab{b}})}]{Saito2001a}
\bibinfo{author}{\bibfnamefont{H.}~\bibnamefont{Saito}} \bibnamefont{and}
  \bibinfo{author}{\bibfnamefont{M.}~\bibnamefont{Ueda}},
  \bibinfo{journal}{Phys. Rev. A} \textbf{\bibinfo{volume}{63}},
  \bibinfo{pages}{043601} (\bibinfo{year}{2001}{\natexlab{b}}),
  \urlprefix\url{https://link.aps.org/doi/10.1103/PhysRevA.63.043601}.

\bibitem[{\citenamefont{Donley et~al.}(2001)\citenamefont{Donley, Claussen,
  Cornish, Roberts, Cornell, and Wieman}}]{Donley2001}
\bibinfo{author}{\bibfnamefont{E.~A.} \bibnamefont{Donley}},
  \bibinfo{author}{\bibfnamefont{N.~R.} \bibnamefont{Claussen}},
  \bibinfo{author}{\bibfnamefont{S.~L.} \bibnamefont{Cornish}},
  \bibinfo{author}{\bibfnamefont{J.~L.} \bibnamefont{Roberts}},
  \bibinfo{author}{\bibfnamefont{E.~A.} \bibnamefont{Cornell}},
  \bibnamefont{and} \bibinfo{author}{\bibfnamefont{C.~E.}
  \bibnamefont{Wieman}}, \bibinfo{journal}{Nature}
  \textbf{\bibinfo{volume}{412}}, \bibinfo{pages}{295} (\bibinfo{year}{2001}).

\bibitem[{\citenamefont{Altin et~al.}(2011)\citenamefont{Altin, Dennis,
  McDonald, D\"oring, Debs, Close, Savage, and Robins}}]{Altin2011}
\bibinfo{author}{\bibfnamefont{P.~A.} \bibnamefont{Altin}},
  \bibinfo{author}{\bibfnamefont{G.~R.} \bibnamefont{Dennis}},
  \bibinfo{author}{\bibfnamefont{G.~D.} \bibnamefont{McDonald}},
  \bibinfo{author}{\bibfnamefont{D.}~\bibnamefont{D\"oring}},
  \bibinfo{author}{\bibfnamefont{J.~E.} \bibnamefont{Debs}},
  \bibinfo{author}{\bibfnamefont{J.~D.} \bibnamefont{Close}},
  \bibinfo{author}{\bibfnamefont{C.~M.} \bibnamefont{Savage}},
  \bibnamefont{and} \bibinfo{author}{\bibfnamefont{N.~P.}
  \bibnamefont{Robins}}, \bibinfo{journal}{Phys. Rev. A}
  \textbf{\bibinfo{volume}{84}}, \bibinfo{pages}{033632}
  (\bibinfo{year}{2011}),
  \urlprefix\url{https://link.aps.org/doi/10.1103/PhysRevA.84.033632}.

\bibitem[{\citenamefont{Claussen et~al.}(2003)\citenamefont{Claussen,
  Kokkelmans, Thompson, Donley, Hodby, and Wieman}}]{Claussen2003}
\bibinfo{author}{\bibfnamefont{N.~R.} \bibnamefont{Claussen}},
  \bibinfo{author}{\bibfnamefont{S.~J. J. M.~F.} \bibnamefont{Kokkelmans}},
  \bibinfo{author}{\bibfnamefont{S.~T.} \bibnamefont{Thompson}},
  \bibinfo{author}{\bibfnamefont{E.~A.} \bibnamefont{Donley}},
  \bibinfo{author}{\bibfnamefont{E.}~\bibnamefont{Hodby}}, \bibnamefont{and}
  \bibinfo{author}{\bibfnamefont{C.~E.} \bibnamefont{Wieman}},
  \bibinfo{journal}{Phys. Rev. A} \textbf{\bibinfo{volume}{67}},
  \bibinfo{pages}{060701} (\bibinfo{year}{2003}),
  \urlprefix\url{https://link.aps.org/doi/10.1103/PhysRevA.67.060701}.

\bibitem[{\citenamefont{Kuhn et~al.}(2014)\citenamefont{Kuhn, McDonald,
  Hardman, Bennetts, Everitt, Altin, Debs, Close, and Robins}}]{Kuhn2014}
\bibinfo{author}{\bibfnamefont{C.~C.~N.} \bibnamefont{Kuhn}},
  \bibinfo{author}{\bibfnamefont{G.~D.} \bibnamefont{McDonald}},
  \bibinfo{author}{\bibfnamefont{K.~S.} \bibnamefont{Hardman}},
  \bibinfo{author}{\bibfnamefont{S.}~\bibnamefont{Bennetts}},
  \bibinfo{author}{\bibfnamefont{P.~J.} \bibnamefont{Everitt}},
  \bibinfo{author}{\bibfnamefont{P.~A.} \bibnamefont{Altin}},
  \bibinfo{author}{\bibfnamefont{J.~E.} \bibnamefont{Debs}},
  \bibinfo{author}{\bibfnamefont{J.~D.} \bibnamefont{Close}}, \bibnamefont{and}
  \bibinfo{author}{\bibfnamefont{N.~P.} \bibnamefont{Robins}},
  \bibinfo{journal}{New Journal of Physics} \textbf{\bibinfo{volume}{16}},
  \bibinfo{pages}{073035} (\bibinfo{year}{2014}),
  \urlprefix\url{http://stacks.iop.org/1367-2630/16/i=7/a=073035}.

\bibitem[{\citenamefont{Sachdev}(2011)}]{Sachdev2011}
\bibinfo{author}{\bibfnamefont{S.}~\bibnamefont{Sachdev}},
  \emph{\bibinfo{title}{{Quantum Phase Transitions}}}
  (\bibinfo{publisher}{{Cambridge University Press}},
  \bibinfo{address}{{Cambridge}}, \bibinfo{year}{2011}).

\bibitem[{\citenamefont{Dziarmaga}(2010)}]{Dziarmaga2010}
\bibinfo{author}{\bibfnamefont{J.}~\bibnamefont{Dziarmaga}},
  \bibinfo{journal}{Advances in Physics} \textbf{\bibinfo{volume}{59}},
  \bibinfo{pages}{1063} (\bibinfo{year}{2010}),
  \eprint{https://doi.org/10.1080/00018732.2010.514702},
  \urlprefix\url{https://doi.org/10.1080/00018732.2010.514702}.

\bibitem[{\citenamefont{Anderson}(1958)}]{Anderson1958}
\bibinfo{author}{\bibfnamefont{P.~W.} \bibnamefont{Anderson}},
  \bibinfo{journal}{Phys. Rev.} \textbf{\bibinfo{volume}{109}},
  \bibinfo{pages}{1492} (\bibinfo{year}{1958}),
  \urlprefix\url{http://link.aps.org/doi/10.1103/PhysRev.109.1492}.

\bibitem[{\citenamefont{M\"uller and
  Delande}(2011)}]{MuellerDelande:Houches:2009}
\bibinfo{author}{\bibfnamefont{C.~A.} \bibnamefont{M\"uller}} \bibnamefont{and}
  \bibinfo{author}{\bibfnamefont{D.}~\bibnamefont{Delande}},
  \emph{\bibinfo{title}{{Disorder and interference: localization phenomena}}}
  (\bibinfo{publisher}{Oxford Scholarship}, \bibinfo{year}{2011}),
  vol.~\bibinfo{volume}{91}, chap.~\bibinfo{chapter}{9},
  \eprint{arXiv:1005.0915}.

\bibitem[{\citenamefont{Kosior et~al.}(2015)\citenamefont{Kosior, Major,
  P\l{}odzie\ifmmode~\acute{n}\else \'{n}\fi{}, and Zakrzewski}}]{Kosior2015}
\bibinfo{author}{\bibfnamefont{A.}~\bibnamefont{Kosior}},
  \bibinfo{author}{\bibfnamefont{J.}~\bibnamefont{Major}},
  \bibinfo{author}{\bibfnamefont{M.}~\bibnamefont{P\l{}odzie\ifmmode~\acute{n}\else
  \'{n}\fi{}}}, \bibnamefont{and}
  \bibinfo{author}{\bibfnamefont{J.}~\bibnamefont{Zakrzewski}},
  \bibinfo{journal}{Phys. Rev. A} \textbf{\bibinfo{volume}{92}},
  \bibinfo{pages}{023606} (\bibinfo{year}{2015}),
  \urlprefix\url{https://link.aps.org/doi/10.1103/PhysRevA.92.023606}.

\bibitem[{\citenamefont{Major}(2016)}]{Major2016}
\bibinfo{author}{\bibfnamefont{J.}~\bibnamefont{Major}},
  \bibinfo{journal}{Phys. Rev. A} \textbf{\bibinfo{volume}{94}},
  \bibinfo{pages}{053613} (\bibinfo{year}{2016}),
  \urlprefix\url{https://link.aps.org/doi/10.1103/PhysRevA.94.053613}.

\bibitem[{\citenamefont{Major et~al.}(2017)\citenamefont{Major,
  P\l{}odzie\ifmmode~\acute{n}\else \'{n}\fi{}, Dutta, and
  Zakrzewski}}]{Major2017}
\bibinfo{author}{\bibfnamefont{J.}~\bibnamefont{Major}},
  \bibinfo{author}{\bibfnamefont{M.}~\bibnamefont{P\l{}odzie\ifmmode~\acute{n}\else
  \'{n}\fi{}}}, \bibinfo{author}{\bibfnamefont{O.}~\bibnamefont{Dutta}},
  \bibnamefont{and}
  \bibinfo{author}{\bibfnamefont{J.}~\bibnamefont{Zakrzewski}},
  \bibinfo{journal}{Phys. Rev. A} \textbf{\bibinfo{volume}{96}},
  \bibinfo{pages}{033620} (\bibinfo{year}{2017}),
  \urlprefix\url{https://link.aps.org/doi/10.1103/PhysRevA.96.033620}.

\bibitem[{\citenamefont{Fishman et~al.}(1982)\citenamefont{Fishman, Grempel,
  and Prange}}]{Fishman:LocDynAnders:PRL82}
\bibinfo{author}{\bibfnamefont{S.}~\bibnamefont{Fishman}},
  \bibinfo{author}{\bibfnamefont{D.~R.} \bibnamefont{Grempel}},
  \bibnamefont{and} \bibinfo{author}{\bibfnamefont{R.~E.}
  \bibnamefont{Prange}}, \bibinfo{journal}{Phys. Rev. Lett.}
  \textbf{\bibinfo{volume}{49}}, \bibinfo{pages}{509} (\bibinfo{year}{1982}),
  \urlprefix\url{link.aps.org/doi/10.1103/PhysRevLett.49.509}.

\bibitem[{\citenamefont{Casati et~al.}(1989)\citenamefont{Casati, Guarneri, and
  Shepelyansky}}]{Casati:IncommFreqsQKR:PRL89}
\bibinfo{author}{\bibfnamefont{G.}~\bibnamefont{Casati}},
  \bibinfo{author}{\bibfnamefont{I.}~\bibnamefont{Guarneri}}, \bibnamefont{and}
  \bibinfo{author}{\bibfnamefont{D.~L.} \bibnamefont{Shepelyansky}},
  \bibinfo{journal}{Phys. Rev. Lett.} \textbf{\bibinfo{volume}{62}},
  \bibinfo{pages}{345} (\bibinfo{year}{1989}),
  \urlprefix\url{link.aps.org/doi/10.1103/PhysRevLett.62.345}.

\bibitem[{\citenamefont{Lemari\'e et~al.}(2009)\citenamefont{Lemari\'e,
  Chab\'e, Szriftgiser, Garreau, Gr\'emaud, and
  Delande}}]{Lemarie:Anderson3D:PRA09}
\bibinfo{author}{\bibfnamefont{G.}~\bibnamefont{Lemari\'e}},
  \bibinfo{author}{\bibfnamefont{J.}~\bibnamefont{Chab\'e}},
  \bibinfo{author}{\bibfnamefont{P.}~\bibnamefont{Szriftgiser}},
  \bibinfo{author}{\bibfnamefont{J.~C.} \bibnamefont{Garreau}},
  \bibinfo{author}{\bibfnamefont{B.}~\bibnamefont{Gr\'emaud}},
  \bibnamefont{and} \bibinfo{author}{\bibfnamefont{D.}~\bibnamefont{Delande}},
  \bibinfo{journal}{Phys. Rev. A} \textbf{\bibinfo{volume}{80}},
  \bibinfo{pages}{043626} (\bibinfo{year}{2009}), \eprint{0907.3411},
  \urlprefix\url{http://link.aps.org/doi/10.1103/PhysRevA.80.043626}.

\bibitem[{\citenamefont{Sanchez-Palencia
  et~al.}(2007)\citenamefont{Sanchez-Palencia, Cl\'ement, Lugan, Bouyer,
  Shlyapnikov, and Aspect}}]{Sanchez2007}
\bibinfo{author}{\bibfnamefont{L.}~\bibnamefont{Sanchez-Palencia}},
  \bibinfo{author}{\bibfnamefont{D.}~\bibnamefont{Cl\'ement}},
  \bibinfo{author}{\bibfnamefont{P.}~\bibnamefont{Lugan}},
  \bibinfo{author}{\bibfnamefont{P.}~\bibnamefont{Bouyer}},
  \bibinfo{author}{\bibfnamefont{G.~V.} \bibnamefont{Shlyapnikov}},
  \bibnamefont{and} \bibinfo{author}{\bibfnamefont{A.}~\bibnamefont{Aspect}},
  \bibinfo{journal}{Phys. Rev. Lett.} \textbf{\bibinfo{volume}{98}},
  \bibinfo{pages}{210401} (\bibinfo{year}{2007}),
  \urlprefix\url{https://link.aps.org/doi/10.1103/PhysRevLett.98.210401}.

\bibitem[{\citenamefont{Billy et~al.}(2008)\citenamefont{Billy, Josse, Zuo,
  Bernard, Hambrecht, Lugan, Cl\'ement, Sanchez-Palencia, Bouyer, and
  Aspect}}]{Billy2008}
\bibinfo{author}{\bibfnamefont{J.}~\bibnamefont{Billy}},
  \bibinfo{author}{\bibfnamefont{V.}~\bibnamefont{Josse}},
  \bibinfo{author}{\bibfnamefont{Z.}~\bibnamefont{Zuo}},
  \bibinfo{author}{\bibfnamefont{A.}~\bibnamefont{Bernard}},
  \bibinfo{author}{\bibfnamefont{B.}~\bibnamefont{Hambrecht}},
  \bibinfo{author}{\bibfnamefont{P.}~\bibnamefont{Lugan}},
  \bibinfo{author}{\bibfnamefont{D.}~\bibnamefont{Cl\'ement}},
  \bibinfo{author}{\bibfnamefont{L.}~\bibnamefont{Sanchez-Palencia}},
  \bibinfo{author}{\bibfnamefont{P.}~\bibnamefont{Bouyer}}, \bibnamefont{and}
  \bibinfo{author}{\bibfnamefont{A.}~\bibnamefont{Aspect}},
  \bibinfo{journal}{Nature} \textbf{\bibinfo{volume}{453}},
  \bibinfo{pages}{891} (\bibinfo{year}{2008}).

\bibitem[{\citenamefont{Roati et~al.}(2008)\citenamefont{Roati, D'Errico,
  Fallani, Fattori, Fort, Zaccanti, Modugno, Modugno, and
  Inguscio}}]{Roati2008}
\bibinfo{author}{\bibfnamefont{G.}~\bibnamefont{Roati}},
  \bibinfo{author}{\bibfnamefont{C.}~\bibnamefont{D'Errico}},
  \bibinfo{author}{\bibfnamefont{L.}~\bibnamefont{Fallani}},
  \bibinfo{author}{\bibfnamefont{M.}~\bibnamefont{Fattori}},
  \bibinfo{author}{\bibfnamefont{C.}~\bibnamefont{Fort}},
  \bibinfo{author}{\bibfnamefont{M.}~\bibnamefont{Zaccanti}},
  \bibinfo{author}{\bibfnamefont{G.}~\bibnamefont{Modugno}},
  \bibinfo{author}{\bibfnamefont{M.}~\bibnamefont{Modugno}}, \bibnamefont{and}
  \bibinfo{author}{\bibfnamefont{M.}~\bibnamefont{Inguscio}},
  \bibinfo{journal}{Nature} \textbf{\bibinfo{volume}{453}},
  \bibinfo{pages}{895} (\bibinfo{year}{2008}).

\bibitem[{\citenamefont{Basko et~al.}(2006)\citenamefont{Basko, Aleiner, and
  Altschuler}}]{Basko06}
\bibinfo{author}{\bibfnamefont{D.}~\bibnamefont{Basko}},
  \bibinfo{author}{\bibfnamefont{I.}~\bibnamefont{Aleiner}}, \bibnamefont{and}
  \bibinfo{author}{\bibfnamefont{B.}~\bibnamefont{Altschuler}},
  \bibinfo{journal}{Ann. Phys. (NY)} \textbf{\bibinfo{volume}{321}},
  \bibinfo{pages}{1126} (\bibinfo{year}{2006}).

\bibitem[{\citenamefont{Oganesyan and Huse}(2007)}]{Oganesyan07}
\bibinfo{author}{\bibfnamefont{V.}~\bibnamefont{Oganesyan}} \bibnamefont{and}
  \bibinfo{author}{\bibfnamefont{D.~A.} \bibnamefont{Huse}},
  \bibinfo{journal}{Phys. Rev. B} \textbf{\bibinfo{volume}{75}},
  \bibinfo{pages}{155111} (\bibinfo{year}{2007}),
  \urlprefix\url{http://link.aps.org/doi/10.1103/PhysRevB.75.155111}.

\bibitem[{\citenamefont{\ifmmode \check{Z}\else
  \v{Z}\fi{}nidari\ifmmode~\check{c}\else \v{c}\fi{}
  et~al.}(2008)\citenamefont{\ifmmode \check{Z}\else
  \v{Z}\fi{}nidari\ifmmode~\check{c}\else \v{c}\fi{}, Prosen, and
  Prelov\ifmmode~\check{s}\else \v{s}\fi{}ek}}]{Znidaric08}
\bibinfo{author}{\bibfnamefont{M.}~\bibnamefont{\ifmmode \check{Z}\else
  \v{Z}\fi{}nidari\ifmmode~\check{c}\else \v{c}\fi{}}},
  \bibinfo{author}{\bibfnamefont{T.}~\bibnamefont{Prosen}}, \bibnamefont{and}
  \bibinfo{author}{\bibfnamefont{P.}~\bibnamefont{Prelov\ifmmode~\check{s}\else
  \v{s}\fi{}ek}}, \bibinfo{journal}{Phys. Rev. B}
  \textbf{\bibinfo{volume}{77}}, \bibinfo{pages}{064426}
  (\bibinfo{year}{2008}),
  \urlprefix\url{http://link.aps.org/doi/10.1103/PhysRevB.77.064426}.

\bibitem[{\citenamefont{Huse et~al.}(2014)\citenamefont{Huse, Nandkishore, and
  Oganesyan}}]{Huse14}
\bibinfo{author}{\bibfnamefont{D.~A.} \bibnamefont{Huse}},
  \bibinfo{author}{\bibfnamefont{R.}~\bibnamefont{Nandkishore}},
  \bibnamefont{and}
  \bibinfo{author}{\bibfnamefont{V.}~\bibnamefont{Oganesyan}},
  \bibinfo{journal}{Phys. Rev. B} \textbf{\bibinfo{volume}{90}},
  \bibinfo{pages}{174202} (\bibinfo{year}{2014}),
  \urlprefix\url{http://link.aps.org/doi/10.1103/PhysRevB.90.174202}.

\bibitem[{\citenamefont{Nandkishore and Huse}(2015)}]{Rahul15}
\bibinfo{author}{\bibfnamefont{R.}~\bibnamefont{Nandkishore}} \bibnamefont{and}
  \bibinfo{author}{\bibfnamefont{D.~A.} \bibnamefont{Huse}},
  \bibinfo{journal}{Ann. Rev. Cond. Mat. Phys.} \textbf{\bibinfo{volume}{6}},
  \bibinfo{pages}{15} (\bibinfo{year}{2015}).

\bibitem[{\citenamefont{Kozarzewski et~al.}(2016)\citenamefont{Kozarzewski,
  Prelov\ifmmode~\check{s}\else \v{s}\fi{}ek, and
  Mierzejewski}}]{kozarzewski2016}
\bibinfo{author}{\bibfnamefont{M.}~\bibnamefont{Kozarzewski}},
  \bibinfo{author}{\bibfnamefont{P.}~\bibnamefont{Prelov\ifmmode~\check{s}\else
  \v{s}\fi{}ek}}, \bibnamefont{and}
  \bibinfo{author}{\bibfnamefont{M.}~\bibnamefont{Mierzejewski}},
  \bibinfo{journal}{Phys. Rev. B} \textbf{\bibinfo{volume}{93}},
  \bibinfo{pages}{235151} (\bibinfo{year}{2016}),
  \urlprefix\url{https://link.aps.org/doi/10.1103/PhysRevB.93.235151}.

\bibitem[{\citenamefont{Mierzejewski et~al.}(2016)\citenamefont{Mierzejewski,
  Herbrych, and Prelov\ifmmode~\check{s}\else \v{s}\fi{}ek}}]{mierzejewski2016}
\bibinfo{author}{\bibfnamefont{M.}~\bibnamefont{Mierzejewski}},
  \bibinfo{author}{\bibfnamefont{J.}~\bibnamefont{Herbrych}}, \bibnamefont{and}
  \bibinfo{author}{\bibfnamefont{P.}~\bibnamefont{Prelov\ifmmode~\check{s}\else
  \v{s}\fi{}ek}}, \bibinfo{journal}{Phys. Rev. B}
  \textbf{\bibinfo{volume}{94}}, \bibinfo{pages}{224207}
  (\bibinfo{year}{2016}),
  \urlprefix\url{https://link.aps.org/doi/10.1103/PhysRevB.94.224207}.

\bibitem[{\citenamefont{Sierant and Zakrzewski}(2018)}]{Sierant2018}
\bibinfo{author}{\bibfnamefont{P.}~\bibnamefont{Sierant}} \bibnamefont{and}
  \bibinfo{author}{\bibfnamefont{J.}~\bibnamefont{Zakrzewski}},
  \bibinfo{journal}{New Journal of Physics} \textbf{\bibinfo{volume}{20}},
  \bibinfo{pages}{043032} (\bibinfo{year}{2018}),
  \urlprefix\url{http://stacks.iop.org/1367-2630/20/i=4/a=043032}.

\end{thebibliography}

\end{document}